\documentclass[twocolumn, switch]{article} % Method A for two-column formatting

\usepackage{graphicx}
\usepackage{preprint}

\usepackage{amsmath, amsthm, amssymb, amsfonts}

\usepackage[numbers,square]{natbib}
\usepackage[utf8]{inputenc}	% allow utf-8 input
\usepackage[T1]{fontenc}	% use 8-bit T1 fonts
\usepackage{xcolor}		% colors for hyperlinks
\usepackage[colorlinks = true,
            linkcolor = purple,
            urlcolor  = blue,
            citecolor = cyan,
            anchorcolor = black]{hyperref}	% Color links to references, figures, etc.
\usepackage{booktabs} 		% professional-quality tables
\usepackage{comment}
\usepackage{nicefrac}		% compact symbols for 1/2, etc.
\usepackage{microtype}		% microtypography
\usepackage{lineno}		% Line numbers
\usepackage{float}			% Allows for figures within multicol

\usepackage{algorithmic}
\usepackage{algorithm}
\usepackage{array}

\usepackage{newfloat}
\DeclareFloatingEnvironment[name={Supplementary Figure}]{suppfigure}
\usepackage{sidecap}
\sidecaptionvpos{figure}{c}

\usepackage{titlesec}
\titlespacing\section{0pt}{12pt plus 3pt minus 3pt}{1pt plus 1pt minus 1pt}
\titlespacing\subsection{0pt}{10pt plus 3pt minus 3pt}{1pt plus 1pt minus 1pt}
\titlespacing\subsubsection{0pt}{8pt plus 3pt minus 3pt}{1pt plus 1pt minus 1pt}

\usepackage{tikz,xcolor,hyperref}

\usepackage[labelformat=simple]{subcaption}

\usepackage{textcomp}
\usepackage{stfloats}
\usepackage{url}
\usepackage{verbatim}
\usepackage{calc}
\usepackage{xspace}
\newcommand{\approachName}{FPTP\xspace}
\newcommand{\mdbver}{7.0.1}
\newcommand{\relname}{MongoDB \mdbver\xspace}
\newcommand{\af}[1]{{\bf AF: #1}}

\usepackage[english]{babel}
\usepackage{listings}
\definecolor{lime}{HTML}{A6CE39}
\DeclareRobustCommand{\orcidicon}{
	\begin{tikzpicture}
	\draw[lime, fill=lime] (0,0) 
	circle [radius=0.16] 
	node[white] {{\fontfamily{qag}\selectfont \tiny ID}};
	\draw[white, fill=white] (-0.0625,0.095) 
	circle [radius=0.007];
	\end{tikzpicture}
	\hspace{-2mm}
}
\foreach \x in {A, ..., Z}{\expandafter\xdef\csname orcid\x\endcsname{\noexpand\href{https://orcid.org/\csname orcidauthor\x\endcsname}
			{\noexpand\orcidicon}}
}
\title{First~Past~the~Post: Evaluating~Query~Optimization~in~MongoDB}

\usepackage{authblk}

\author[1]{Dawei Tao}
\author[1]{Enqi Liu}
\author[2]{Sidath Randeni Kadupitige\orcidC}
\author[2\thanks{Some work completed while M. Cahill was employed by MongoDB.}]{Michael Cahill\orcidD}
\author[2]{Alan Fekete\orcidE}
\author[2]{Uwe R{\"o}hm\orcidF}

\affil[1,2]{School of Computer Science, University of Sydney, Sydney, Australia.}
\affil[1]{\texttt{\{dtao7193, eliu5850\}@uni.sydney.edu.au}}
\affil[2]{\texttt{\{sidath.randenikadupitige, michael.cahill, alan.fekete, uwe.roehm\}@sydney.edu.au} \newline
Research was supported by Australian Research Council Linkage Project grant LP160100883}

\begin{document}

\twocolumn[ % Method A for two-column formatting
    \begin{@twocolumnfalse} % Method A for two-column formatting
      
    \maketitle
    
    % Note: TKDE requires abstracts of less than 200 words.

\begin{abstract}
Query optimization is crucial for every database management system (DBMS) to enable fast execution of declarative queries.
Most DBMS designs include cost-based query optimization. However, MongoDB implements a different approach to choose an execution plan that we call \emph{``first past the post'' (\approachName) query optimization}.  \approachName does not estimate costs for each execution plan, but rather partially executes the alternative plans in a round-robin race and observes the work done by each relative to the number of records returned.
%The optimizer computes a score for each potential plan based on the partial execution and then chooses the plan with the highest score to run to completion for the query.

 %One of our initial steps has been to create a generalized query model for evaluating the effectiveness of the query optimizer.
In this paper, we analyze the effectiveness of MongoDB's \approachName query optimizer. We see whether the optimizer chooses the best execution plan among the alternatives and measure how the chosen plan compares to the optimal plan. We also show how to visualize the effectiveness and identify situations where the \relname query optimizer chooses suboptimal query plans. Through experiments, we conclude that \approachName has
%what we call
a preference bias, choosing index scans even in many cases where collection scans would run faster. 
We identify the reasons for the preference bias, which can lead MongoDB to choose a plan with more than twice the runtime compared to the optimal plan for the query.

\end{abstract}

    \vspace{0.35cm}
    \end{@twocolumnfalse} % Method A for two-column formatting
]

\section{Introduction}
Query optimization is a long-established topic in database management systems. For a given declarative query submitted by a user, there are many possible execution plans, each of which describes a correct way to calculate the result. The plans for a query can vary in cost by orders of magnitude, so a query optimizer is vital to choose an efficient plan among the possibilities. Most database management systems include an optimizer that is cost-based: it considers a variety of plans, estimates the cost of each plan from statistics, knowledge of the index structures, etc., and then chooses to execute the plan with lowest estimated cost among those it considered. 

MongoDB~\cite{mongodb_2019} is a popular document-oriented DBMS with a very different approach to query optimization which is not based on estimating the costs of queries before they are run. Instead, MongoDB  runs many execution plans in a round-robin ``race'', allowing each to do a small amount of work at a time. After a point, it considers the progress of each plan and calculates a score based on the number of results produced for the work done to that point. The plan with the highest score wins the race, and it alone runs to completion as the chosen plan for this query. We call the MongoDB technique "first past the post" (\approachName) query optimization.  
%As far as we know, no previous research has evaluated the effectiveness of the \approachName approach.

The central aim of our work is to evaluate and understand the implementation of MongoDB's \approachName optimization. In this paper, we explain the query optimization approach of MongoDB in Section~\ref{sec:background}. We describe our innovative approach to evaluation of query optimization (and how we visualize the results)
in Section~\ref{sec:methodology}.
The results of our empirical study of MongoDB are presented in Section~\ref{sec:evaluation}. We find that \approachName in MongoDB has a preference bias: it systematically avoids choosing collection scans even for queries where they are the best plan. We explore the reasons for this in Section~\ref{sec:rootcauseanalysis}.

This paper makes the following contributions:
\begin{enumerate}
    \item We describe in detail how the \approachName query optimizer in MongoDB chooses query plans.
    \item We propose an innovative way to evaluate and visualize the impact on query performance of an optimizer's choices. 
    \item By using this approach, we identify places where the MongoDB query optimizer chooses suboptimal query plans.
    Our approach could form the basis of an automated regression testing tool to verify that the query planner in MongoDB improves over time.
    \item We identify causes of the preference bias we found in \approachName, in which index scans are systematically chosen even when a collection scan runs faster.
\end{enumerate}

Source code relating to this work is available from: \href{https://github.com/michaelcahill/mongodb-fptp-paper}{github.com/michaelcahill/mongodb-fptp-paper}

\section{Background and related work}
\label{sec:background}
There exists an extensive literature on query
optimization. Here, we point to some of the fundamental papers and then describe MongoDB (the system we study in this paper) and how its \approachName query optimization is performed.

%%%%%%
%% Query Optimization Concepts
%%%%%%%%%%%%%%%%%%%%%%

\subsection{Query Optimization Architecture}

A vital factor in the spread of relational database technology was the support for declarative queries, in which a user expresses a query indicating {\it what} information is needed and the system then discovers {\it how to} execute the query, that is, what steps to take to retrieve that information from what is stored~\cite{Codd70}. Declarative queries have remained important also with other data models, such as object and document data. In general, a given query will have many execution plans (the details will depend on physical structures such as indexes), and these plans will differ greatly in performance. For declarative queries to be workable, the database management system must be able to optimize, that is, find an execution approach which is not only correct in retrieving the specified information, but also runs rapidly.

Although some early commercial systems used pure heuristics to choose the execution plan for a submitted query, the dominant approach is cost-based optimization, invented by Selinger and colleagues for System~R~\cite{SelingerACLP79}. The essential work of a cost-based optimizer is (i) generate a variety of \emph{candidates}, which are execution plans each of which would calculate the correct results for the given query, (ii) estimate the cost each candidate plan will incur when it runs, and (iii) choose to actually execute the plan, among those considered, whose estimate cost is lowest. Cost-based optimization is now industry standard practice~\cite{lahdenmaki2005relational}.

Research has continued apace, both in industry and academia, to improve the design of query optimizers.

\subsection{Candidate Plan Generation} \label{sec:candidate-plan-generation}
For complicated queries, there are a very large number of execution plans that produce the correct results. The plans can vary logically (for example, in which order joins are done, whether selection happens before or after a join, etc.) and physically (e.g., which index is used to find records, which join algorithm is used). The System~R optimizer~\cite{SelingerACLP79} restricted its attention to joins performed in a particular set of sequential patterns. The generation of potential logical plans by successively applying transformations was introduced by Freytag~\cite{Freytag87} and Graefe~\cite{GraefeD87} and used in IBM's Starburst project~\cite{HaasFLP89, PiraheshHH92}.  Later, Graefe extended this to a unified view in which logical and physical choices were treated together to generate potential plans~\cite{Graefe95a}.

\subsection{Cost Estimation}
The cost one estimates is a measure of the resources consumed in calculation; traditionally, for a disk-based database management system, the dominant cost is bringing pages from disk to memory or vice versa. To estimate the cost of an execution plan, one proceeds from estimates of the cost of each operation (e.g. join, select) within the plan. The cost of a step depends on the physical structure and especially on the cardinality of the inputs and outputs of the step. For base collections, cardinality is clear, but subsequent operations take as inputs the results of prior calculations, and so estimating the cardinality requires estimating the selectivity of predicates, the number of distinct values in an attribute, the probability that items will match in a join, etc. Many techniques have been proposed for estimating specific operations such as projection, range, join, and aggregation~\cite{ahad1989estimating, HaasNSS96, PoosalaIHS96, MarklMKTHS05}. Estimates are often based on statistics kept in the database or calculated dynamically by sampling~\cite{olken1995random}.

The cost estimate for a plan is very sensitive to errors in the cardinality estimates, so several researchers have tried to achieve \emph{robust} planning, where the plan choice works well even if the estimates are mistaken. One approach is to include a measure of uncertainty in each estimate~\cite{babcock2005towards}, and then choose a plan that is good throughout the likely range of estimates. Another is to check while the chosen plan is executing and to stop and re-optimize if the reality differs too much from what was estimated~\cite{markl2004robust}. These ideas have been combined~\cite{babu2005proactive}.

\subsection{Optimizer Implementation and Use}
Graefe and others have emphasized the importance of modularity and extensibility of the optimizer~\cite{Graefe94, Graefe95a}. An important theme has been to allow parallel execution in the optimizer~\cite{WaasH09, SolimanAREGSCGRPWNKB14}. 
Chaudhuri and colleagues at Microsoft and Lohman and others at IBM pioneered the use of the cost estimator component as a tool to recommend better physical structures, e.g., deciding which indexes to create~\cite{Chaudhuri:1998478, Chaudhuri:20049cd, Valentin:20005d5, Agrawal:2005dff, chaudhuri2008pay}.

Recent work has focused on learning from dynamic query execution for adaptive code generation to avoid adding complex, specialized query execution techniques to support diverse hardware~\cite{Gubner22}; and using reinforcement learning to adaptively change query plans during execution~\cite{Trummer21}. MongoDB's \approachName approach was designed before this work, and while it is simpler, it has been in widespread use for many years and thus is still of interest.

\subsection{Evaluating Query Optimization}
While much of the literature has taken an engineering approach, designing better ways to do optimization, there has also been work that is scientific in style, looking at how to evaluate the quality of a given optimizer. Our paper follows this tradition. For cost-based optimizers, there has been study of how accurate the estimates are, by comparing the estimated cost with the measured cost when a plan is run~\cite{GuSW12, LeisGMBK015}. This does not of course apply to MongoDB, since \approachName does not use cost estimates. Another aspect of evaluation is to find a suitable set of queries on which to test the optimizer. In this paper we use simple conjunctive queries with two range predicates, but more complex queries are important in practice, and some systems have generated random queries with a mix of predicates and joins, eg~\cite{StillgerF95, WaasG00}. Our work has been most influenced by the work of Haritsa and colleagues, whose focus is on understanding which plan is chosen for each query, especially as selectivity varies. The Picasso tool~\cite{reddy2005analyzing, Haritsa10} introduced plan diagram visulization, which is one of the displays we use here. This led to a new goal for an optimizer: to be stable (that is, to have a plan that works well not just for the given query, but also for nearby ones)~\cite{AbhiramaBDSH10}.

%%%%%%
%% MongoDB 
%%%%%%%%%%%%%%%%%%%%%%
\subsection{Document Data Model}
MongoDB is a document-oriented database designed for ease of development and scaling. A MongoDB database stores collections of documents. MongoDB documents (the equivalent of records in a relational DBMS) do not need to have a schema defined beforehand \cite{chandra2015base, mongodb_2019}. Instead, the fields in each document can be chosen on the fly.  This flexible design allows developers to represent hierarchical relationships, to store arrays, and other more complex structures simply.

% [Michael] I'm inclined not to over-sell.
% The  design of MongoDB environments are very scalable. Companies across the world have deployed clusters running more than a hundred nodes with around millions of documents within the database.

The key components of the MongoDB data model design are the database, collection, document, and cursor \cite{banker2011mongodb}.

% [Michael] I think we could cut this table for space without losing much
\begin{comment}
Table \ref{table:terms} shows the relationship of RDBMS terminology with MongoDB. 

\begin{table}[htb]
    \begin{tabular}{ll}
        \toprule
        MongoDB    &    RDBMS\\
        \midrule
        Database    &      Database\\
        Collection   &      Table\\
        Document      &     Record(s)\\
        Field          &      Column\\
        Embedded Documents &  Foreign Key\\
%         Primary Key (key \_id) & Primary Key\\
        \bottomrule
    \end{tabular}
    
    \caption{MongoDB terminology versus RDBMS terminology.}
    \label{table:terms}
\end{table}
\end{comment}

\subsection{MongoDB Query Model}

\begin{figure}[t]
  \centering
  \includegraphics[width=0.9\linewidth]{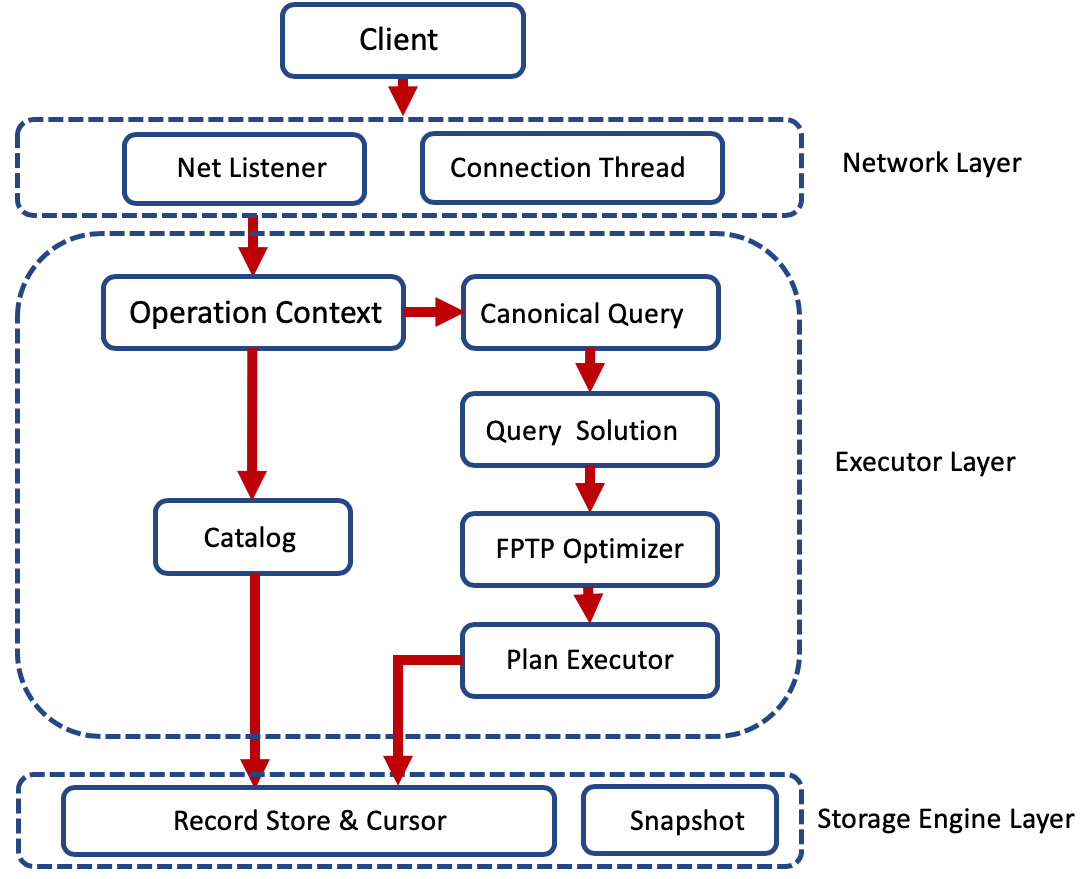}
  \caption{MongoDB query processing workflow.}
  \label{fig:workflow}
\end{figure}

At a low level, MongoDB stores data as BSON documents, which extends the JSON model to offer more data types and efficient encoding and decoding. APIs in a variety of programming languages make it easy for programmers to create those documents. This coupled with the similarity between MongoDB's JSON-like document model and the data structures used in object-oriented programming, makes integration with applications simple. 

MongoDB applications can use complex queries, secondary B+-Tree indexes, and aggregations to retrieve unstructured, semi-structured, and structured data. A vital factor in this flexibility is MongoDB's support for various query types. Queries can return documents, document projections, or complex aggregations calculated over
many documents \cite{mongodb_2019}.

% [Michael] this introduces range queries but not sure whether it's otherwise needed
\begin{comment}
\begin{itemize}
     \item Key-value queries: find documents where a field, usually the primary key, matches a given value.
     \item Range queries: find documents where field values match inequality predicates (e.g, greater than, less than or equal to, between).
     \item Text search: find documents in relevance order based on text arguments using Boolean operators such as AND, OR, NOT.
\end{itemize}
\end{comment}

Here, we have a closer look at MongoDB's \verb|find()| query command and introduce
how it relates to SQL SELECT expressions.
%Example \ref{alg:findsyntex} shows the syntax of MongoDB \verb|find()| command.
\begin{lstlisting}[caption={MongoDB find() query example}, label={alg:queryexample}]
db.runCommand({
    "find": "movie",
    "filter": { 
        movieid: {"$gte": 0, "$lte":1000}, 
        avgrating: {"$gte": 5},
    },
    "projection": {moviename: 1, avgrating: 1},
    "sort": {avgrating, -1},
    "limit": 10,
})
\end{lstlisting}

The most common operators for a \verb|find()| command are \verb|filter| (analogous to the SQL WHERE expression, with an implicit AND between conditions), \verb|projection| (analogous to the SQL SELECT expression) and \verb|sort|. MongoDB query is different from SQL in many ways: the query is expressed in BSON format. Each operator in the query language corresponds to a field in a document that can be included or omitted as needed.

\subsection{Execution Plan Generation and Caching}

\begin{figure}[tb]
  \centering
  \includegraphics[width=0.75\linewidth]{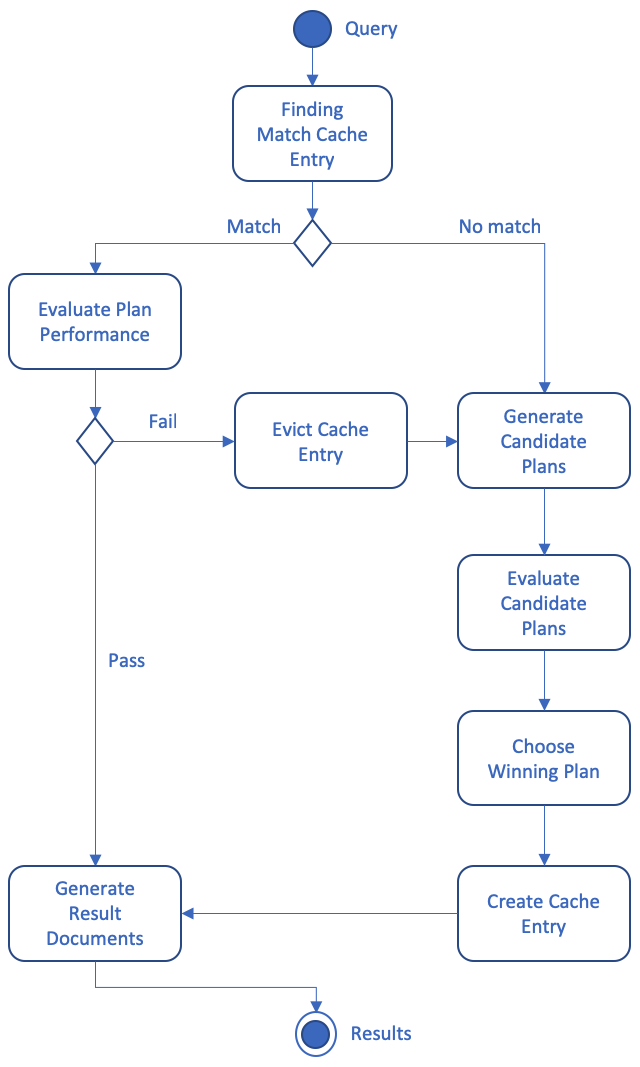}
  \caption{Logical flow of MongoDB query optimizer.}
  \label{figure:logicflow}
\end{figure}

The design of the MongoDB query processor and optimizer is evolving with each release. We provide detailed insight into the state-of-the-art architecture. The overall design of the optimizer is clean and orderly. However, in later sections we demonstrate that there is room for further improvement.

In this section, we explain how a MongoDB query is submitted, parsed, and optimized before it interacts with the MongoDB storage engine. Explanations in this section are mainly gathered from interviews with professional MongoDB developers and from examining MongoDB source code.

Figure \ref{fig:workflow} provides an overview of the query processing workflow. We break down the whole process into three layers. In the network layer, MongoDB specifies a MongoDB Wire Protocol which is a simple socket-based, request-response-style protocol. Clients connect to the database following the protocol. In the executor layer, queries are received in the form of operation contexts. Some queries such as \verb|insert()| do not require optimization. Therefore, these queries interact directly with the storage engine.

Queries that require optimization are standardized and simplified to Canonical Queries ignoring values in the predicate, which MongoDB calls the \textit{query shape}:\footnote{\url{https://mongodb.com/docs/manual/reference/glossary/\#std-term-query-shape}}. 
The query shape is designed for the cache mechanism; generation and evaluation of query plans uses a complete query with concrete values. 

\begin{comment}
``A combination of query predicate, sort, projection, and collation. The query shape allows MongoDB to identify logically equivalent queries and analyze their performance.
For the query predicate, only the structure of the predicate, including the field names, are significant. The values in the query predicate are insignificant. Therefore, a query predicate \verb|{ type: 'food' }| is equivalent to the query predicate \verb|{ type: 'utensil' }| for a query shape.''

The process of generating a canonical query mainly involves match expression optimization. A match expression is an expression consisting of logical operations in the \verb|filter| operator. To optimize a match expression, MongoDB extracts all logical operators and constitutes an expression tree.  In this process, MongoDB optimizes the ordering of logical expressions and eliminates duplicates. The simplified expression tree can be used to filter records later. 
\end{comment}

For a given Canonical Query, the Query Solution module might yield multiple execution plans. Each execution plan is a tree of query stages, where the leaf stages access data via a collection or index. When there are multiple relevant indexes, execution plans are generated to explore the combinations. Multiple execution plans will also be generated when stages can be reordered or there are multiple implementations of a given stage (e.g., \verb|$or| queries and joins expressed with \verb|$lookup|).

All candidate execution plans are evaluated by the \approachName query optimizer. Finally, the candidate plan assigned the best efficiency score by \approachName is executed by the plan executor. The plan is also cached, and if the same query shape is executed soon after, the cached plan will be reused as long as its efficiency remains below 10 times the evaluated efficiency (with default settings).

In this work, we mainly focus on investigating the query optimizer's query plan evaluation strategy but not the efficiency of its cache mechanism. In the experimental analysis, we programmatically force MongoDB not to use the query plan cache. 

\subsection{MongoDB's \approachName Query Optimizer}

\begin{comment} Michael's notes
For future references, \ref{alg:es} is summarising the code in src/mongo/db/exec/multi_plan.cpp, specifically the main loops in MultiPlanStage::pickBestPlan and MultiPlanStage::workAllPlans.
\end{comment}

\begin{algorithm}[tb]
    \caption{Algorithm used for early termination of the race}
    \begin{algorithmic}
        \REQUIRE $collection$, the main data source for the query, and $candidates$, a list of candidate plans
        \ENSURE statistics describing each plan's trial execution
        \STATE $N \gets numRecords(collection)$
        \STATE $maxWorks \gets max(10000, 0.3N)$
        \STATE $maxResults \gets 101$ \COMMENT{with default configuration}
        \STATE $working \gets$ \TRUE
        \STATE $i \gets 0$
        \STATE $plan.results \gets 0$ $\forall$ $plan \in candidates$
        \STATE
        \WHILE{$working$ \AND $i < maxWorks$}
            \FORALL{$plan \in candidates$}
                \STATE $state \gets plan.work()$
                \IF{$state$ = ADVANCED}
                    \STATE $plan.results \gets plan.results + 1$
                    \IF{$plan.results \ge maxResults$}
                        \STATE $working \gets$ \FALSE
                    \ENDIF
                \ELSIF{$state$ = EOF}
                    \STATE $working \gets$ \FALSE
                \ENDIF
            \ENDFOR
            \STATE $i \gets i + 1$
        \ENDWHILE
    \end{algorithmic}
    \label{alg:es}
\end{algorithm}

We now describe the mechanism of MongoDB's query optimizer. A simplified explanation of \approachName is that the query optimizer initially executes all potential query plans in parallel and chooses the first one that completes a predefined amount of work. The winner of the race is then run to completion as the execution plan and is also cached for future queries with the same shape. 

The \approachName approach consists of two key steps: measuring all candidate plans in a race and then making a decision, that is, choosing the most efficient query plan based on the measurements. During the race phase, all candidate query plans are executed in a round-robin fashion. That is, MongoDB starts all candidate query plans and cycles through them to step through the first parts of the execution of the query using each plan. During the race, time slices are assigned to each query plan in circular order, handling all potential plans. Each time slice performs one logical unit of work representing one step of a query plan, such as examining the next document and checking the predicates on it. Meanwhile, the query optimizer gathers execution metrics and then provides a score for each plan. 

The query optimizer asks each plan for the next document, via a call to a \verb|work()| function. If the step can provide a document, it responds with \verb|ADVANCED|. If all documents from the plan have been retrieved, \verb|work()| returns \verb|EOF| and \verb|working| is set to \verb|false|. However, the given query could be expensive, so MongoDB specifies limits to terminate the race early, as described in Algorithm \ref{alg:es}.

The race stops if the maximum allowable amount of work has been reached, or the requested number of documents has been retrieved (as reflected in the counter $plan.results$), or a plan has completed and returned all its results. The configuration knob $internal\-Query\-Plan\-Evaluation\-Works$ (default 10,000) is the maximum allowed amount of work. However, for very large collections, MongoDB takes a fraction of the number of documents to determine the maximum allowable amount of work. The configuration knob $internal\-Query\-Plan\-Evaluation\-Max\-Results$ (default 0.3) is the maximum number of documents that can be retrieved during the race.

\begin{table}[htb]
    \begin{tabular}{ll}
        \toprule
        Metric         & Value\\
        \midrule
        baseScore      & 1                                           \\
        Productivity   & queryResults / workUnits                    \\
        TieBreak       & min(1.0 / (10 * workUnits), 1e-4)           \\
        noFetchBonus   & TieBreak or 0                               \\
        noSortBonus    & TieBreak or 0                               \\
        noIxisectBonus & TieBreak or 0                               \\
        tieBreakers    & noFetchBonus + noSortBonus + \\
                       & noIxisectBonus \\
        eofBonus       & 0 or 1\\
        \bottomrule
    \end{tabular}
    \caption{MongoDB Query Plan Performance Metrics}
    \label{table:pm}
\end{table}

When the race is terminated, each candidate query plan is assigned a performance score based on the performance metrics shown in Table~\ref{table:pm}. The query optimizer chooses the query plan that has the highest performance score as the execution plan. The performance score is calculated as:
\begin{align*}
 score =~ &baseScore + productivity~+ \\
         &tieBreakers + eofBonus
\end{align*}

Each query plan has a base score of 1. The $productivity$ of a plan is measured as the number of results returned divided by the total amount of work performed. $tieBreakers$ is a very small bonus number that is given when a plan contains no fetch operation ($noFetchBonus$), no blocking sort ($noSortBonus$) or avoids index intersection ($noIxisectBonus$). $eofBonus$ is given if all possible documents are retrieved during the execution of the plan.
\section{Evaluation Methodology}% Effectiveness of \approachName}
\label{sec:methodology}
We now describe a detailed methodology to experimentally judge how well a query optimizer chooses query plans. We describe this methodology as we use it to assess the effectiveness of MongoDB's \approachName query optimizer --- but note that this approach is not limited to \approachName query optimization, or even to document store optimisation, as it makes no assumptions about the internal workings of the query optimizer. We also describe the visual presentations we use to present the results. The essence of our evaluation is to take queries, and find the actual cost of the plan chosen by MongoDB's \approachName optimizer, and then force the system to use each other possible query plan, and measure the actual cost with that plan. We can thus determine whether \approachName chooses the truly optimal query plan, and also we can quantify the drop in performance between the true optimum and what is chosen by \approachName. In this paper, we  consider queries that all have a very simple query shape, in which a single collection is filtered based on the conjunction of two a range predicate on each of two fields. Depending on which indices exist, there are several potential query plans, including a collection scan, and perhaps one or more using an index scan. The actual cost of running a plan will also depend on the selectivity of the predicates, and perhaps the distribution of data values for each field. Our evaluation thoroughly explores the space of different selectivity levels of the fields. We calculate two numeric metrics: what is the fraction of queries where \approachName chooses the true optimum,  and also an average for the performance impact between the chosen plan and the true best, over this space of queries.

We consider a very simple category of queries in our evaluation. If an optimzer works well on these, one would want to extend the evaluation to more sophisticated queries, such as aggregations, disjunctions, or cross-collection lookups (corresponding, in MongoDB, to relational joins). However, as we show, even for the very simple queries, \approachName often chooses a plan that is far from optimal. So, for the evaluation in this paper, we do not see a need to check on more complex queries.  

%Through this experiment, we find that the query optimizer does not consider collection scan as a candidate query plan; even when it does, it sticks on index scan while collection scan is actually a better choice.  In the next section, we focus on reveal the impact of this issue, and explore the underlying mechanism of \approachName to identify the root cause. 

\subsection{Testing Setup}
The testing environment adopts a typical client-server architecture. A single client communicates with a \relname server that is deployed on an Amazon EC2 (Elastic Compute Cloud) instance. The instance type we chose is \verb|m6i.large|. M6 instances work well with Amazon EBS (Amazon Elastic Block Store) gp3 (general purpose SSD) volumes for instance block storage.
%\begin{comment}
Gp3 is the default EBS volume type for Amazon EC2 instances and provides baseline performance of 3,000 IOPS per volume, regardless of volume size, to meet the performance needs of most applications. 100GB of storage is sufficient for storing our database and metadata. During experiments, we need to make sure that RAM is not the bottleneck.
%\end{comment}
%Since our focus is on how well \approachName chooses query plans, we ensure that our indexes fit entirely in RAM so that the system does not read from disk during query execution. 
The instance type we use has 16GB RAM, which is more than enough to keep all indexes in memory for the workloads we used during testing.

We wrote a Python client application to run the experiments and visualize the results. The client runs on the same instance as the server because the client is idle while queries are running, and the network latency is not relevant for our tests. This evaluation client executes each query and gathers execution statistics from the database using MongoDB's query API.
%\begin{comment}
After the testing is complete, this client also analyzes statistical information and visualizes the results.
%\end{comment}

%\subsection{MongoDB Setup, Database Content, Queries}
\subsection{MongoDB Setup and Database Content}
The command line interface and the configuration file provide MongoDB administrators with many options and settings to control the operation of the database system. 
%We modify the default configuration file \verb|mongod.conf| to  allow  remote  access,  since our  Python  application  runs  on  the client side.
During the experiment, we use the \verb|cursor.explain()| method with \verb|allPlansExecution| mode to inspect the candidate query plans and execution statistics. 

The experiments we show here use one collection that contains $1 \times 10^5$ documents. Each document has two fields, namely A and B. A positive integer in the range $[0, 1\times 10^5)$ is assigned to each field. Different experiments use different choices for the index structures. For example, in the first experiment of Section~\ref{sec:evaluation}, 
% and 
we create indexes A\textunderscore1 and B\textunderscore1 on fields A and B, respectively. In the particular results we report here, we always have fields A and B both with uniform distribution and no repetitions among the documents (ie with $10^5$ distinct values). We also ran some experiments with other distributions for the field values, but we saw the same general trends as for uniform distribution.

Despite the simplicity of this data model, the experimental setup exercises the core of MongoDB's query optimizer. As described in Section~\ref{sec:candidate-plan-generation}, MongoDB generates candidate query plans based on the access methods available for the documents required by the query. With a few exceptions, which are not relevant in these experiments, the optimiser does not consider reordering query stages or multiple join strategies. This keeps the set of candidate plans small, but means that queries must be constructed carefully to perform well\footnote{\url{https://www.red-gate.com/simple-talk/blogs/enjoying-joins-in-mongodb/}}. Further, while we focus on simple queries involving two fields so that the visualizations correspond directly to the data, this approach can be applied to more generally.

\subsection{Query Template}
In order to assess the effectiveness of the query optimizer, we use conjunctive range queries of the following form:
$$ \sigma [ \mathit{lowA} \le A < \mathit{highA} \wedge \mathit{lowB} \le B < \mathit{highB} ] ( \mathit{Collection} ) $$

These are selection queries with two range predicates that allow us to have detailed control over the query selectivity. All the queries we apply have the same shape, finding all documents whose A field lies within a given range, and also the B field value lies in another range. The ranges are set by parameters $\mathit{lowA}$, $\mathit{highA}$, $\mathit{lowB}$, $\mathit{highB}$ whose values determine the selectivities of a particular query of this shape. The queries are then executed using MongoDB's query language, as shown in Algorithm~\ref{algo:queryshape}.

\begin{lstlisting}[caption=Query Shape in MongoDB Query Syntax, label=algo:queryshape]
db.collection.find({
    "A" : {"$gte" : lowA, "$lt" : highA},
    "B" : {"$gte" : lowB, "$lt" : highB}
})
\end{lstlisting}

We can determine the selectivity for range predicate on field $X$, $e_X$, 
using the following formula:
\begin{align*}
\begin{split}
 e_X= \frac{\textit{number of documents in range 
 on field X}}{\textit{total number of documents}} \\
 \end{split}
\end{align*}

\subsection{Methodology}
For each experiment, we create a single client and an Amazon EC2 instance with \relname installed. We then build a database with the appropriate data distribution and create the indexes for the particular experiment.

Then, the client performs queries that have different selectivity on A and B. More specifically, each query initiated by the client is a conjunction of one range predicate on A and the other on B. For each query, the client records the selectivity for the range predicates on field A \& B, the execution plan, and the time taken to execute the plan. 

The client generates random queries of the given shape, picking random values for the parameters, to circumvent any potential workload prediction strategies by the query optimizer.
For each query, the client determines and tracks the selectivity of the predicate on field A, and the selectivity of the predicate on field B.    
The query is submitted to MongoDB, and the optimizer determines the query plan using \approachName and executes the chosen plan. We disabled query plan caching for these experiments, so that MongoDB repeats optimization for every submitted query. The client uses the \verb|explain| operation to determine which MongoDB execution plan chose for the query.

We also determine the runtime cost of each candidate query plan, i.e., which is really the best (and thus we will see whether or not MongoDB is choosing wisely). %As we mentioned in Section~\ref{sec:hint}, 
MongoDB provides users with a \verb|hint(<queryPlan>)| method to customize execution plan selection. We use this method to force the query optimizer to execute each suitable plan. We run each plan ten times; when doing this, we extract the \texttt{execution\-Stats.\-execution\-Time\-Millis} field to retrieve
the wall-clock time in milliseconds required for each query execution. We then average the execution times for a plan and identify the optimal query plan: the one with the lowest average execution time among the possible plans for this query.

\paragraph{\textbf{Outlier handling.}} When we gathered execution times of all query plans, we noted
that there were a few outliers which deviate markedly from other data points. These outliers are
%usually fifty 
many times higher than the mean value. We assume that the likely cause is delays during MongoDB warm-up and unstable performance of our EC2 instances rather than the inherent faults of the benchmarked database. Therefore, we identify these outliers and drop them using the standard 1.5 IQR (interquartile range) rule: We measure the interquartile range  and drop any data point greater than the sum of the third quartile value and 1.5 times IQR, also we drop any measurement less than the first quartile minus 1.5 time IQR.

\paragraph{\textbf{Accuracy and Impact Metrics.}}
We finally assess the effectiveness of a query optimizer by two metrics: 
The measure of the overall success of MongoDB's optimizer is its \emph{accuracy}: the fraction of queries for which the optimizer's chosen plan is actually the best one.

In addition to finding whether or not MongoDB's \approachName optimizer chooses the best plan, we also want to know how much query performance is affected by the choice. If the chosen plan were only slightly slower than the best possible plan, users might find this quite acceptable, but if MongoDB chooses a plan that is much slower than another, this is more serious. So, for each query, we determine the ratio of the runtime of the plan chosen by the optimizer to the runtime of the plan which actually runs fastest for that query. We can consider this for each query, and we also calculate the average over all the queries we use, as the \emph{impact} of the optimizer.

\begin{figure}[thb]
  \centering
  \includegraphics[width=0.7\linewidth]{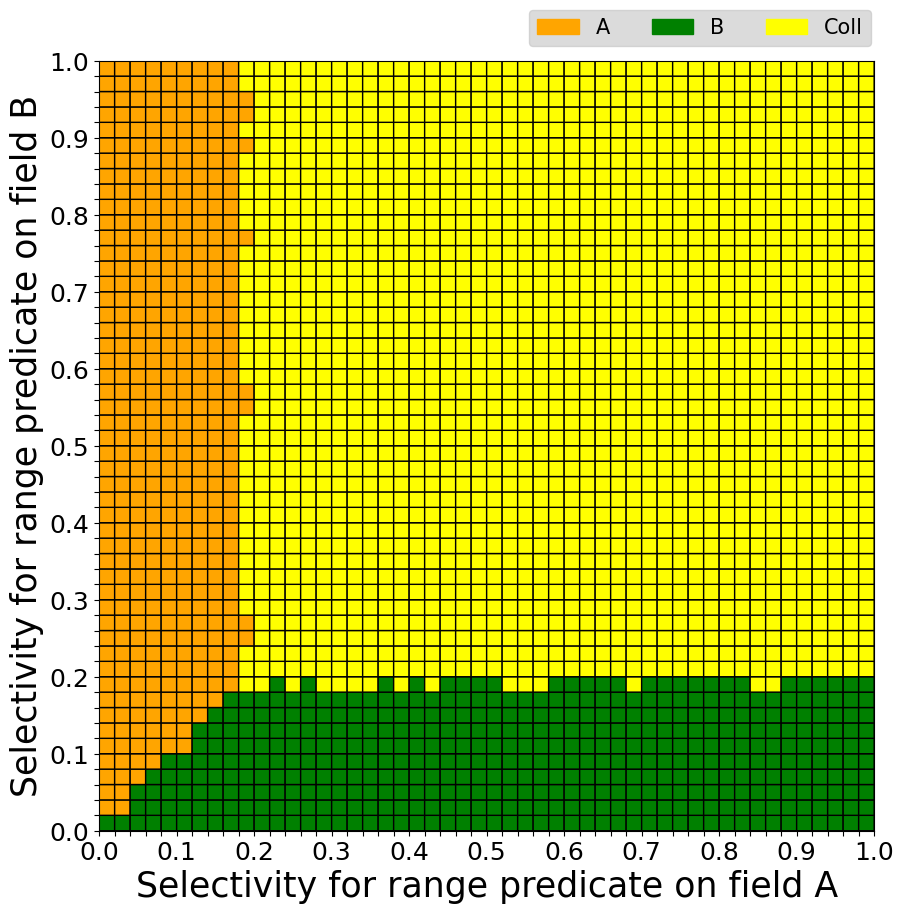}
  \vspace*{-0.5\baselineskip}
  \caption{Visual mapping of execution plans for different selectivities.}
  \label{fig:grid-sample}
  %\Description{\relname}
\end{figure}

\subsection{Visualization of Plan Choices}
\label{sec:vm}

\begin{comment}
\begin{algorithm}[htb]
    \caption{Mapping a pair of selectivities to coordinate}
    \begin{algorithmic}
        
        \STATE $e_A, e_B$ \COMMENT{selectivity for range predicates on A and B}
        \STATE $i, j \in \mathbb{Z}^+$ \COMMENT{column and row number in the grid}
        \STATE $D \gets 50$ \COMMENT{dimension of the grid.}
        \STATE $\delta_e \gets  D*{\frac{100}{D}}$ \COMMENT{step of our measurement}
        \STATE $i \gets e_A / \delta_e$
        \STATE $j \gets e_B / \delta_e$
        
    \end{algorithmic}
    \label{alg:coor}
\end{algorithm}
\end{comment}

\begin{algorithm}[htb]
    \caption{Algorithm used for visual mapping}
    \label{alg:vm}
    \begin{algorithmic}
        \STATE $D \gets $ Dimension of the grid 
        \STATE $M_{p} \gets D \times D$ array of -1 \COMMENT{chosen plan color}
        \STATE $M_{t} \gets D \times D$ array of -1 \COMMENT{plan execution time}
        \STATE $M_{visited} \gets D \times D$ array of 0 \COMMENT{locations visited}
        \STATE $v \gets 0$  \COMMENT{Number of locations visited}
        \STATE $N \gets D^2$  \COMMENT{total number of locations}
        \STATE $a_{min}, a_{max} \gets$  min, max value of field A
        \STATE $b_{min}, b_{max} \gets$  min, max value of field B
        \STATE $e_A, e_B $ \COMMENT{selectivity for range predicate on A and B}
        \STATE $t_p$ \COMMENT{time of seeking execution plan}
        \STATE $P$ \COMMENT{execution plan}
        \STATE $Q$ \COMMENT{MongoDB query}
        \STATE
        \WHILE{$v < N$}
            \STATE $a_{lower}, a_{upper} \gets$ randRangePredicate($a_{min}, a_{max}$)
            \STATE $b_{lower}, b_{upper} \gets$ randRangePredicate($b_{min}, b_{max}$)
            \STATE $e_A \gets$ computeSelectivity($a_{lower}, a_{upper}$)
            \STATE $e_B \gets$ computeSelectivity($b_{lower}, b_{upper}$)
            \STATE $i, j \gets$ mapSelectivityToCoordinate($e_A, e_B$)
            \STATE 
            \IF{$M_{visited}[j][i] = 1$}
                \STATE continue
            \ENDIF
            \STATE
            \STATE $Q \gets$ generateQuery($a_{lower}, a_{upper}, b_{lower}, b_{upper}$)
            \STATE $P, t_p\gets$  db.collection.find($Q$).explain()
            \STATE $M_{p}[j][i] \gets$ colorOfPlan(P)
%            \IF{$P$ is index scan on A}
%                \STATE $M_{p}[j][i] \gets$ white
%            \ELSIF{$P$ is index scan on B}
%                \STATE $M_{p}[j][i] \gets$ black
%            \ELSIF{$P$ is collection scan}
%                 \STATE $M_{p}[j][i] \gets$ yellow
%            \ENDIF
            \STATE $M_{t}[j][i] \gets t_p$ 
            \STATE $M_{visited}[j][i] \gets 1$
            \STATE $v \gets v + 1$
        \ENDWHILE        
    \end{algorithmic}
    \label{alg:map}
\end{algorithm}

We visualize the results of our query optimizer evaluation using two types of visual graphs. The first visual display technique is inspired by the ``plan diagrams'' used by the Picasso system \cite{reddy2005analyzing}. As in Picasso, we consider a family of conjunctive queries with two range predicates, which are parameterised by the selectivity of two attributes. For each choice of selectivities, we identify which query plan is chosen by the optimizer, and plot that decision on a square diagram, where the x and y coordinates reflect the selectivities in that query. Each potential plan is indicated by a different color that can be assigned to the point. %
%one of the relevant research on robust query processing \cite{Karthik:2016d2b}. In that paper, Jayant Haritsa introduced an innovative] way to visualize results in a two-dimensional space. The aim of the piece of work is investigating how query plan cost varies with respect to query selectivity. He used five hyperbolic-shaped contours to separate abounding box into multiple regions, such that the cost of the first contour corresponds to the minimum query cost and the last one corresponds to the maximum cost.He drew query plans on the contour according to their selectivity to compare the cost of different query plans. 
We illustrate the plan chosen by MongoDB by the color displayed in a cell of an appropriately discretized square grid.  We use the same style of diagram for the plan actually chosen by the optimizer and also to illustrate which plan would truly be the fastest for the query. In this way, readers can easily visually compare whether the diagrams are similar to each other (indicating that MongoDB usually chooses the truly best plan) or very different. This also makes clear, which situations lead to incorrect choice by the optimizer.

We describe how a choice of plan is displayed in a plan diagram following \cite{reddy2005analyzing}. For each query, the client will record $e_A, e_B$ and $P_{e_A, e_B}$ (i.e. the execution plan for the query with selectivity $e_A$ and $e_B$). The client will map $P_{e_A, e_B}$ to a position (i, j) on the grid. The magnitudes of $i$ and $j$ are proportional to $e_A$ and $e_B$. Different execution plans are represented by pixels with different colors, as demonstrated in Figure~\ref{fig:grid-sample} (which shows the optimal choice of plan from our first set of experiments). In this example, there are only three possible execution plans, so we use three contrasting colors to distinguish them:

\begin{itemize}
     \item IXSCAN\_A: an index scan uses index A\_1 on field A. This execution plan is represented by an orange pixel. 
    \item IXSCAN\_B: an index scan uses index B\_1 on field B. This execution plan is represented by a green pixel. 
    \item COLLSCAN: a collection scan (i.e. a table scan). This execution plan is represented by a yellow pixel.
\end{itemize}

The grid has dimension $50 \times 50$ and starts gray so that we can identify the abnormal behavior of the query optimizer (i.e. timeout or exceptions are thrown). When the experiment starts, the client keeps generating queries and trying to fill the grid with three colors, until all $50^2$ positions on the grid have been visited.
\begin{comment}
Algorithm \ref{alg:coor} describes how we map a pair of selectivity to a position on the grid.     
\end{comment}
Algorithm~\ref{alg:vm} explains the entire visual mapping process.

% ==================================================================
% but as a FLOAT needs to be given before the page starts...  [UR]
% ==================================================================
\begin{figure*}[t]
\newlength\plotheight%
\setlength\plotheight{\heightof{\includegraphics[width=0.30\textwidth]{images/results-without-covering-index/v\mdbver/comprehensive_practical_winner.png}}}
    \begin{subfigure}[t]{0.3\linewidth}
    \includegraphics[height=\plotheight]{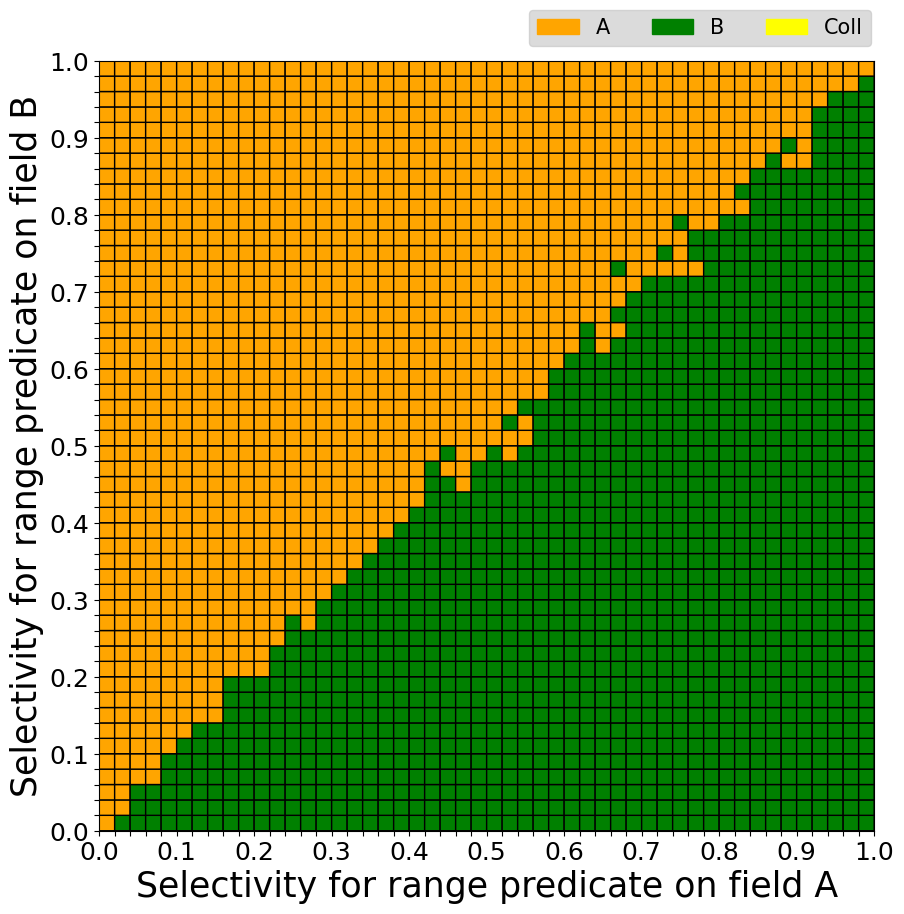}
    \caption{Chosen query plan with both attributes indexed: MongoDB never chooses a collection scan.}
    \label{fig:mongo-bothindexed-choices}
    \end{subfigure}
    \hfill
    \begin{subfigure}[t]{0.3\linewidth}
    \includegraphics[height=\plotheight]{images/results-without-covering-index/v\mdbver/comprehensive_practical_winner.png}
    \caption{Optimal query plan for different selectivities with both attributes indexed.}
    \label{fig:mongo-bothindexed-optimal}
    \end{subfigure}
    \hfill
    \begin{subfigure}[t]{0.3\linewidth}
    \includegraphics[height=\plotheight-1em]{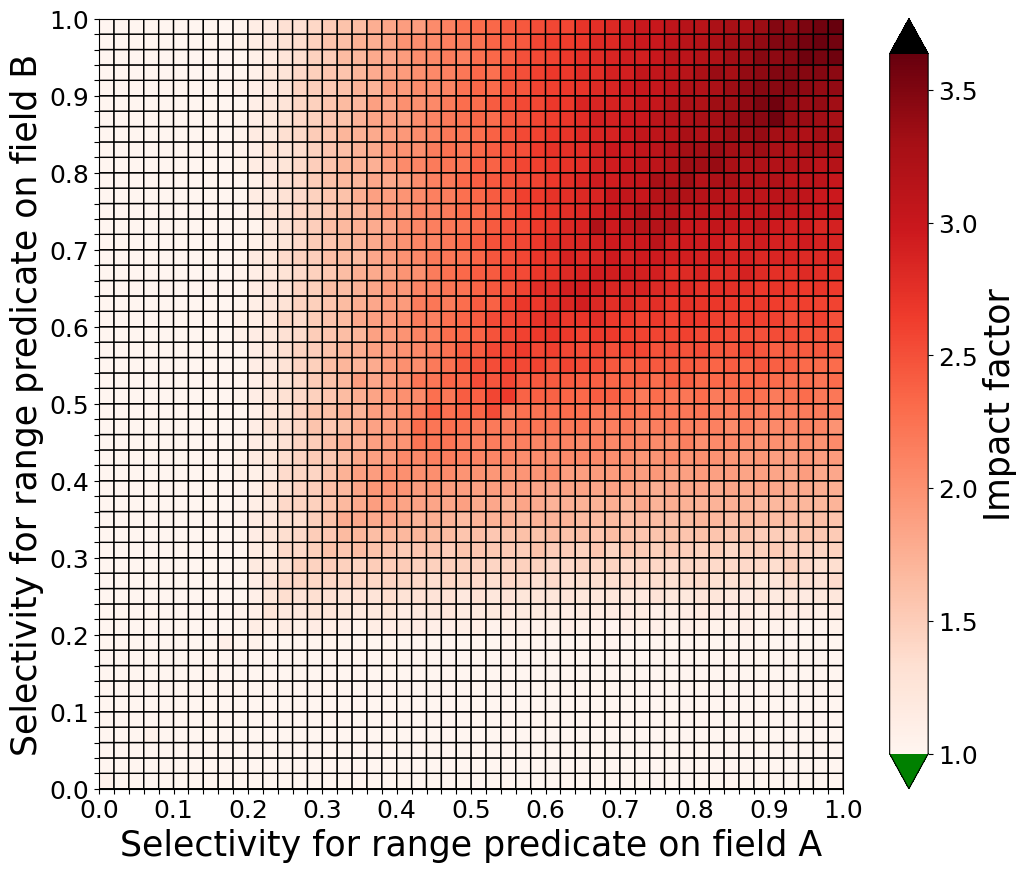}
    \caption{Performance impact of MongoDB's sub-optimal choices.}
    \label{fig:mongo-bothindexed-perfimpact}
    \end{subfigure}
    \caption{Effectiveness of MongoDB's query optimizer with conjunctive filter queries and both attributes indexed.}
    \label{fig:bothindexed-evaluation}
\end{figure*}

\subsection{Visualization of Performance Impact}
We further want to assess how much query performance is impacted by the choices the optimizer makes. To do so, we determine for each query the ratio of the runtime of the plan chosen by the optimizer to the runtime of the plan that actually runs fastest for that query. We propose a heatmap presentation which we believe has not been used before for this aspect of optimizer effectiveness; we plot these ratios on the same discretized grid, with coordinates reflecting the selectivities, and we use a heatmap in which the color saturation corresponds to the ratio between runtime of the chosen plan and runtime of the truly optimal plan for that query.  If the query optimizer has chosen the optimal plan, this ratio will be 1.0 and we represent this with white pixels. 
But the darker the color of a box, the worse the performance impact of the query optimizer's (suboptimal) choice. An example of such a heat map visualizing the impact of \approachName performance is shown in Figure~\ref{fig:mongo-bothindexed-perfimpact}.

\section{Evaluation of MongoDB's \approachName Query Optimizer}
\label{sec:evaluation}

In this section, we use our visualization approach from Section~\ref{sec:methodology} to assess the effectiveness of \approachName query optimization in \relname, in which we see which plans are chosen and which plans would actually be the best. We use varying physical designs for the data (varying sets of indexes).

%We quantify the impact of the performance bias issue and present the results through a heatmap. Through experiments we determine the accuracy of the query optimizer is only 69.29\%. Besides that, the optimal query plan is up to 86.83\% faster than MongoDB's choice. We demonstrate that the overall performance of the MongoDB query optimizer can be improved by 10.96\% if MongoDB adopt the optimal query plans. We then examine various database designs by repeating the experiment on different dataset with various kinds of distribution to further explore the impact of this issue. We find that the distribution of the dataset does not  influence MongoDB's query plan decisions. 

%\subsection{MongoDB Fails to Choose Collection Scans}
\subsection{Querying Indexed Collections}
\label{sec:evaluation_bothindexed}
In our first experiment, we investigate the case of conjunctive filter queries on two attributes where both of the queried attributes are indexed. The physical data design is, as described in Section~\ref{sec:methodology}, a collection with two integer-valued fields (A and B). In this experiment, each field has a uniform data distribution and each field has an index. Therefore, there are three reasonable query execution plans: IXSCAN\_A, IXSCAN\_B and COLLSCAN.
%One is to retrieve the documents that match the predicate on A, using the index on A, and then examine each to see whether the document also meets the predicate on B; this plan is called IXSCAN\_A, an index scan on A. We can symmetrically do an index scan on B using the index on B (IXSCAN\_B). Instead, we can perform a full collection scan, examining each document to see whether it meets the conjunction of both predicates (COLLSCAN). %Therefore, the balanced occurrences of IXSCAN\_A and IXSCAN\_B can be explained by the symmetrical design of the database. 

%We repeat the experiment with the modified version of MongoDB which takes the collection scan into the consideration. Nevertheless, the results remain unchanged, which confirms that MongoDB query optimizer has underlying issues when it evaluates query plans. For convenience,we name the original MongoDB and the modified version  V1 and V2, respectively.

Figure~\ref{fig:mongo-bothindexed-choices} plots the execution plans picked in this case by the \relname query optimizer. We observe that both IXSCAN\_A  and IXSCAN\_B are picked by the optimizer with equal chance, and as expected, the field where the query selectivity is lower (that is, fewer documents satisfy this predicate), has the index used.  This phenomenon suggests that the \approachName approach is capable of making efficient choices between index scans. The boundary between the two index scans is clear and forms a diagonal across the grid. The query optimizer is quite robust: away from the boundary, the chosen plan does not change with small perturbations in the query. 

However, in this experiment, we see that COLLSCAN was never the chosen plan. To see that this is surprising, we present the diagram showing which plan is actually the best for queries in this physical design (in Figure~\ref{fig:mongo-bothindexed-optimal}, which is the same as Figure~\ref{fig:grid-sample}) and what the ratio is between the cost of running the chosen plan compared to the true best plan (Figure~\ref{fig:mongo-bothindexed-perfimpact}). 

%\textbf{[[Diagram update needed: Figure~\ref{fig:mongo-bothindexed-perfimpact} showing the performance impact on two index structure, on \relname unmodified)]]}

\begin{figure*}[t]
%\newlength\plotheight%
\setlength\plotheight{\heightof{\includegraphics[width=0.30\textwidth]{images/results-without-covering-index/v\mdbver/comprehensive_practical_winner.png}}}
    \begin{subfigure}[t]{0.3\linewidth}
    \includegraphics[height=\plotheight]{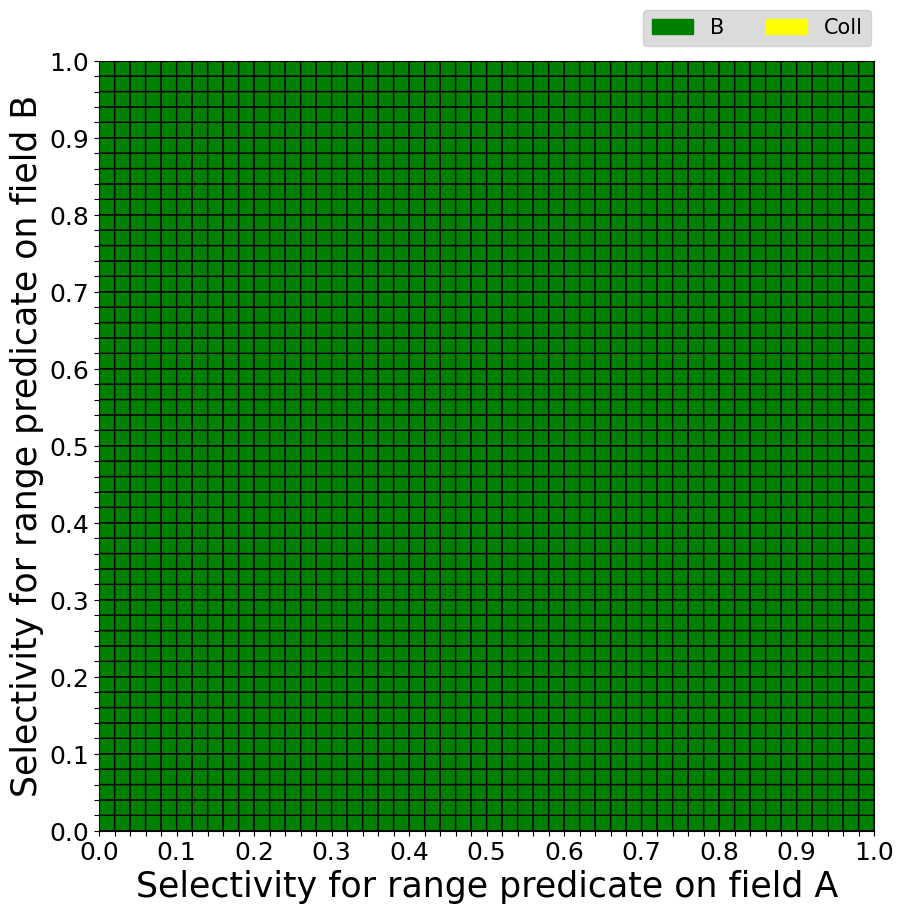}
    \caption{Chosen query plan with only attribute B indexed: MongoDB still never chooses a collection scan.}
    \label{fig:mongo-singleindex-choices}
    \end{subfigure}
    \hfill
    \begin{subfigure}[t]{0.3\linewidth}
    \includegraphics[height=\plotheight]{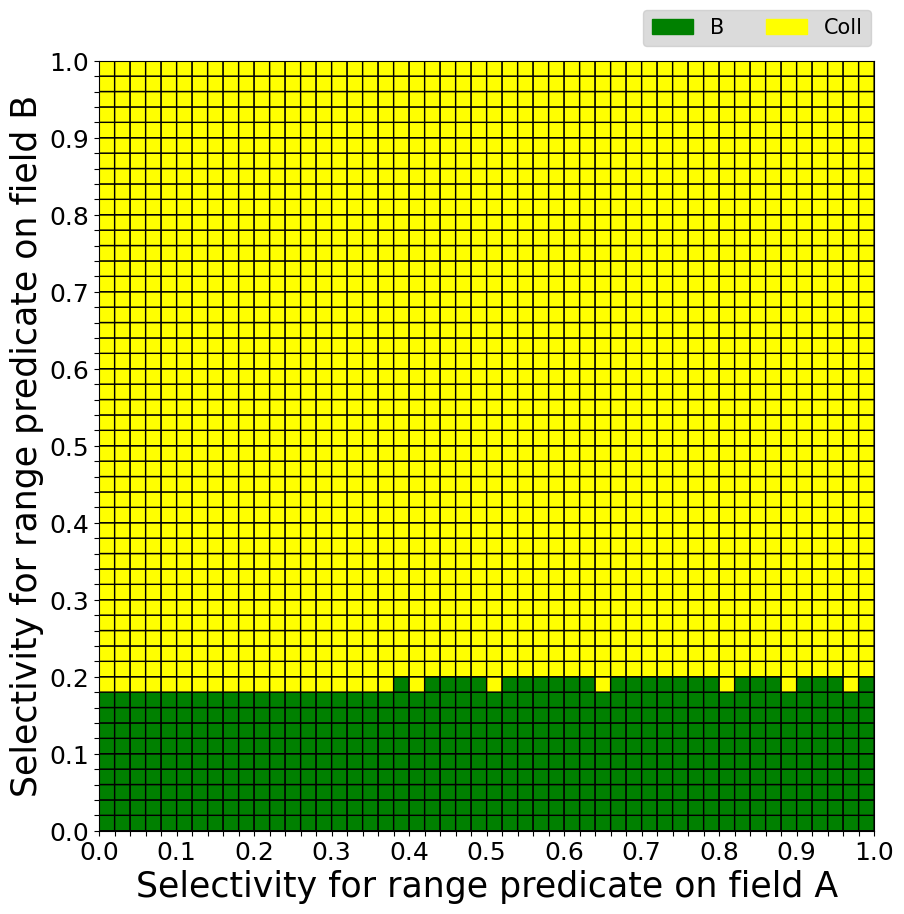}
    \caption{Optimal query plans for different selectivities with only one index.}
    \label{fig:mongo-singleindex-optimal}
    \end{subfigure}
    \hfill
    \begin{subfigure}[t]{0.3\linewidth}
    \includegraphics[height=\plotheight-1em]{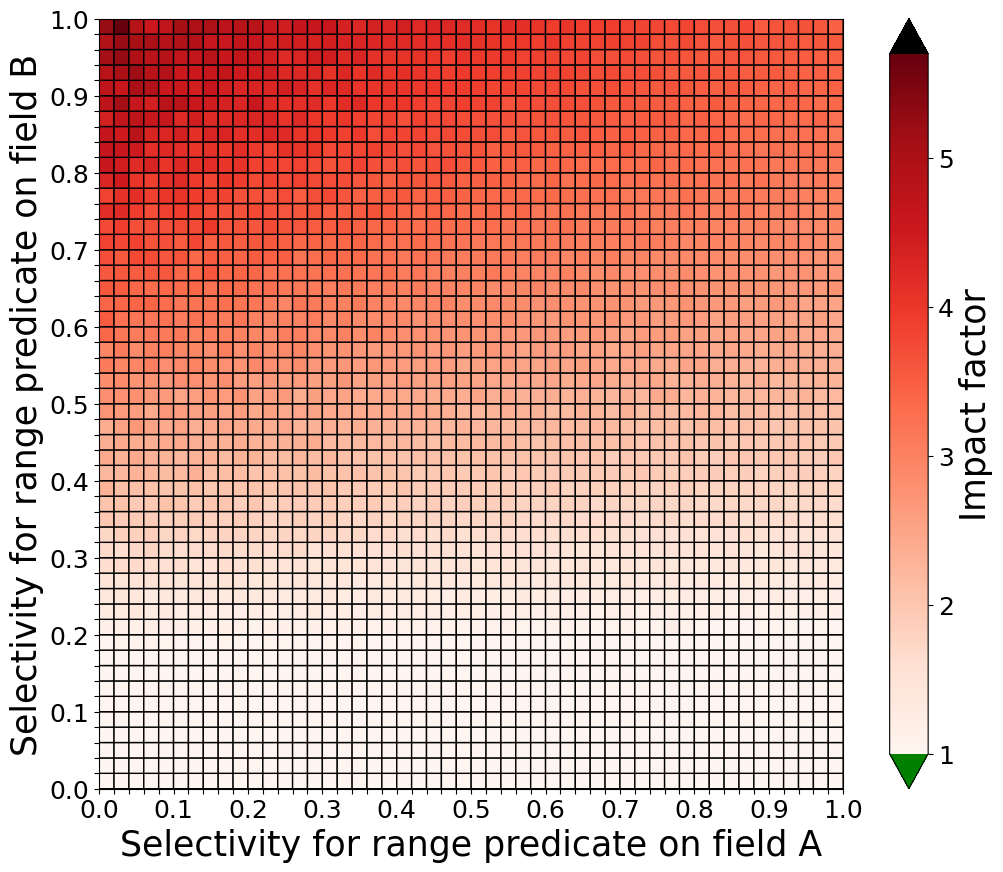}
    \caption{Performance impact of MongoDB's query plan choices.}
    \label{fig:mongo-singleindex-perfimpact}
    \end{subfigure}
    \caption{Effectiveness of MongoDB's query optimizer with conjunctive filter queries and only one attribute indexed.}
    \label{fig:singleindex-evaluation}
\end{figure*}

\begin{figure*}[t]
%\newlength\plotheight%
\setlength\plotheight{\heightof{\includegraphics[width=0.29\textwidth]{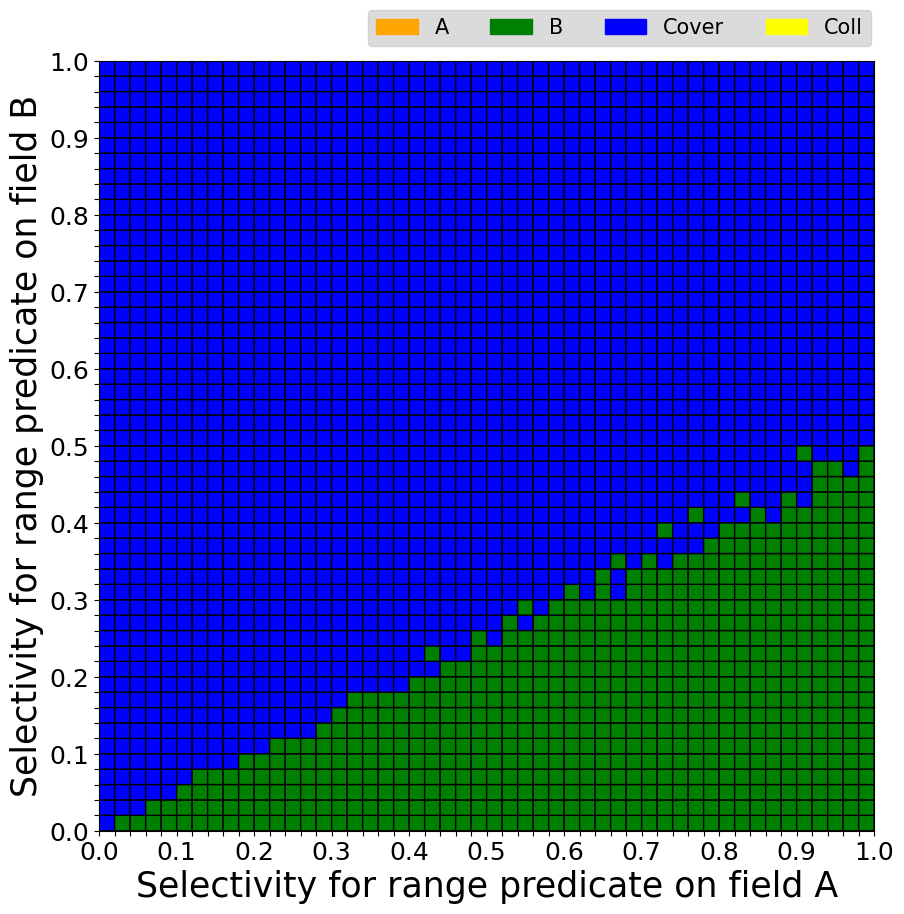}}}
    \begin{subfigure}[t]{0.3\linewidth}
    \includegraphics[height=\plotheight]{images/results-with-covering-index/v\mdbver/comprehensive_mongo_choice.png}
    \caption{Chosen query plan with all attributes indexed and covering index available too.}
    \label{fig:mongo-coveringindex-choices}
    \end{subfigure}
    \hfill
    \begin{subfigure}[t]{0.3\linewidth}
    \includegraphics[height=\plotheight]{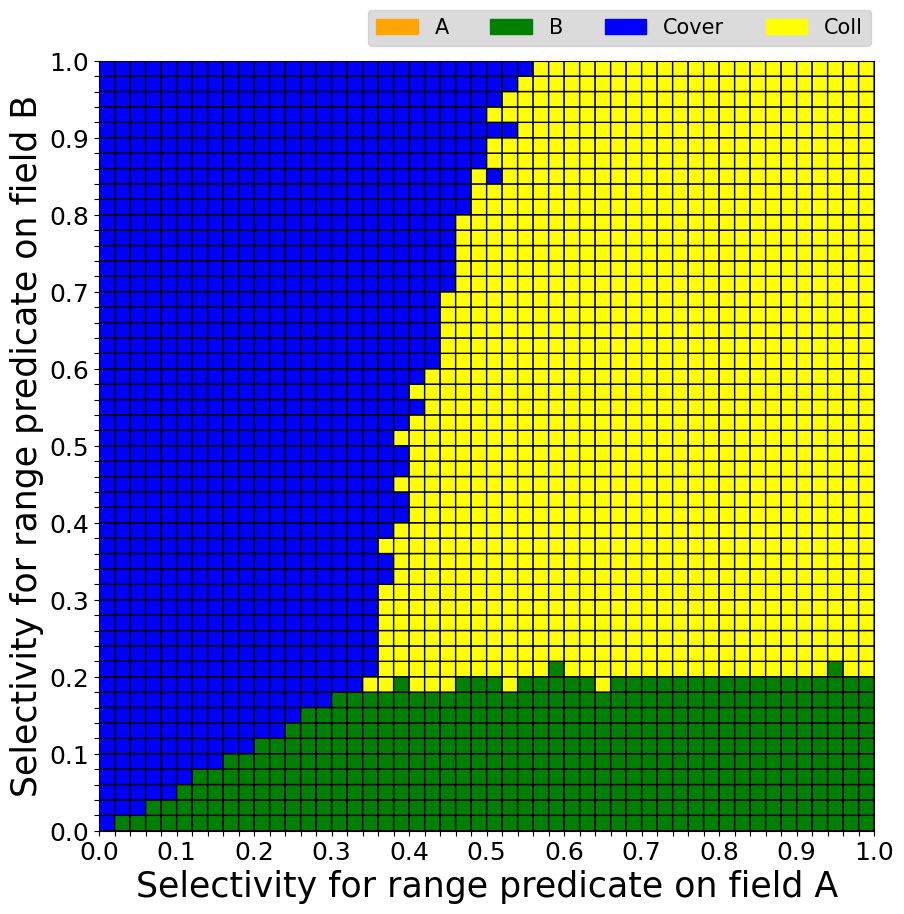}
    \caption{Optimal query plans for different selectivities with covering index available too.}
    \label{fig:mongo-coveringindex-optimal}
    \end{subfigure}
    \hfill
    \begin{subfigure}[t]{0.3\linewidth}
    \includegraphics[height=\plotheight-1em]{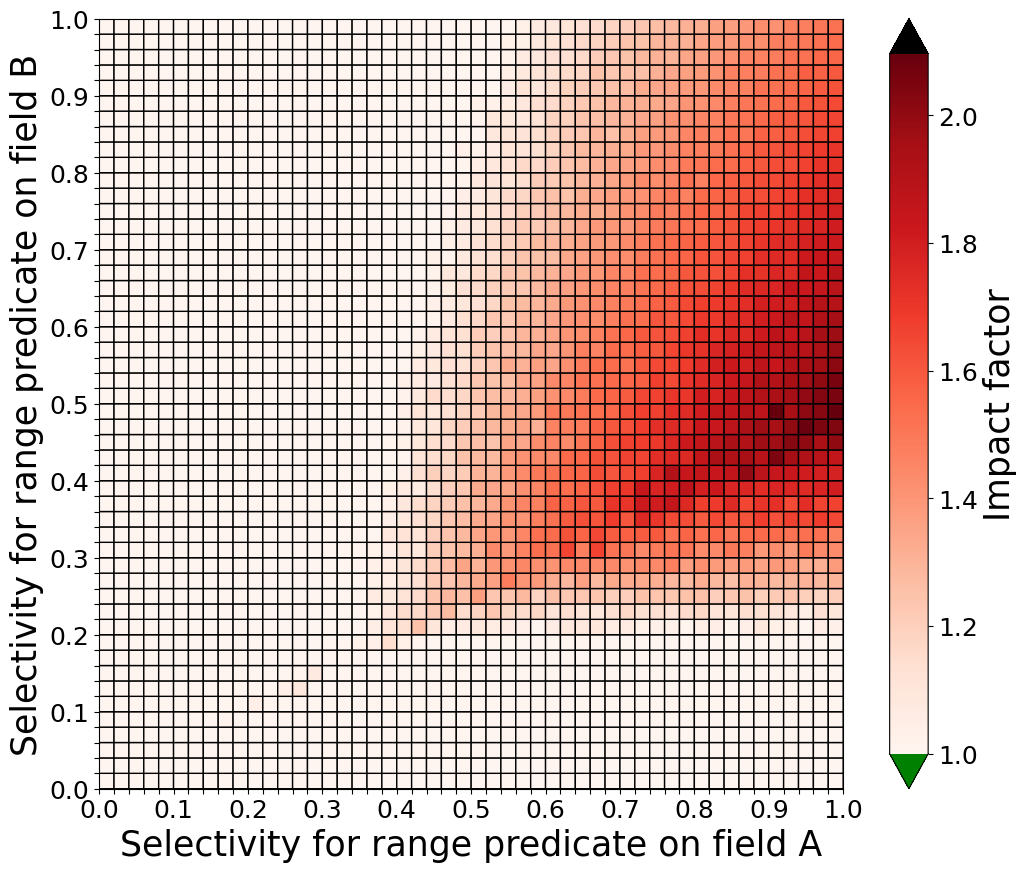}
    \caption{Performance impact of MongoDB's sub-optimal plan choices.}
    \label{fig:mongo-coveringindex-perfimpact}
    \end{subfigure}
    \caption{Effectiveness of MongoDB's query optimizer with all queried attributes being indexed including a covering index.}
    \label{fig:coveringindex-evaluation}
\end{figure*}

As one would expect, the performance of an index scan is significantly worse than a collection scan for a query that has high selectivity on both fields (that is, each predicate is satisfied by many of the documents). At the extreme, if the query retrieves all documents in the collection, an index scan would first retrieve index documents and then use those index information to locate and fetch all documents. In contrast, a collection scan does not have index-access overhead; it directly retrieves all documents without using any index. Therefore, the top right corner of the grid in Figure~\ref{fig:mongo-bothindexed-optimal} is yellow, as expected. Since MongoDB does not choose a collection scan, in this corner of Figure~\ref{fig:mongo-bothindexed-perfimpact}, we see strong red, indicating that the chosen query runs substantially slower than the best possible execution plan.

% Michael: exp1-4, orig, uniform both
In summary, for this experiement, MongoDB's \approachName query optimizer shows an overall accuracy of just shy of 34\%
%(accuracy=33.92)
in this scenario, so about two thirds of the time there would be a better plan than one of the chosen index scan plans. The average impact factor of all those optimizer decisions is 1.7 (on average, the chosen query plan is 70\% slower than the optimal plan), with some chosen plans more than three times slower than a collection scan.

%We perform a case study to investigate the reason why the query optimizer does not choose collection scans. To begin with, we focus on the case where the selectivity of range predicate A and B both equal to one. In this case, the query intends to retrieve all documents in the collection. We execute the query with \verb|explain(allPlanExecution)| method to inspect the query execution statistics.  We observe that COLLSCAN is neither in the \verb|rejectedPlans| nor in the \verb|winningPlan|. We find out that MongoDB V1 does not consider COLLSCAN as a candidate query plan for this case, while COLLSCAN is theoretically the optimal query plan. To verify this observation, we inspect all query plans in the experiment, it turns out COLLSCAN has never been examined by the query optimizer. In conclusion, we prove that the MongoDB V1 has a defect that it does not consider collection scan as a candidate plan through a visualization and a case study. 

\vspace*{-0.5\baselineskip}
\subsection{Single Index Scenario}
%Our previous experiment only covered the scenario in which field A and field B both have an index. 
In the next experiment, we investigate the behavior of MongoDB's query optimizer with a physical structure where there is an index for only one of the attributes (field B). Again, we have uniformly distributed distinct values in each field.

%Table~\ref{tbl:case5-8} shows the physical designs used in this section.

%\begin{table}[h]
%\begin{tabular}{lllll}
%\toprule
%{\small Case} & {\small Distribution of A} & {\small Distribution o%f B} & {\small Index on A} & {\small Index on B} \\ 
%\midrule
%5    & Uniform           & Uniform           & False      & True       \\
%6    & Uniform           & Linear            & False      & True       \\
%7    & Uniform           & Normal            & False      & True       \\
%8    & Uniform           & Zipfian           & False      & True   \\
%bottomrule
%\end{tabular}
%\caption{The database design of case 5, 6, 7 and 8 }
%\label{tbl:case5-8}
%\end{table}

Figure~\ref{fig:mongo-singleindex-choices} plots the query plans chosen by \relname and Figure~\ref{fig:mongo-singleindex-optimal} shows the optimal query plans. Figure~\ref{fig:mongo-singleindex-perfimpact} shows the performance impact of the optimizer's choices. 
%Assuming MongoDB V1: Figure~\ref{fig:mongo-singleindex-choices} indicates that the collection scan has chance  to be chosen after we adding it to the query plan candidates. 
 %Nevertheless, 
%With the visualization in Figure~\ref{fig:mongo-singleindex-perfimpact}, we should observe that collection scan only been chosen when the impactfactor reaches the upper bound. In other words, MongoDB only realizes a collection scan is better when the relative performance of an index scan becomes extremely inefficient. We will analysis the root causeof this phenomenon in Section~\ref{sec:rootcauseanalysis}. %OR: \ref{chap:irc} ???

% Michael: exp2-4, orig, uniform single
From Figures~\ref{fig:mongo-singleindex-choices} and \ref{fig:mongo-singleindex-optimal} we see that the collection scan is actually the best as long as more than about 20\% of records satisfy the range predicate on B; however, even in these situations MongoDB chooses to use the available index on B; again, it never does a collection scan. Thus, the optimizer's choice is even less accurate than in the previous scenario, with an overall accuracy of 19\%. %(19.0\%)

And these sub-optimal choices sometimes come with a huge cost. Near the boundary between optimal choices, the two have similar costs (and so there is not much impact from choosing the index scan); however, in cases where most documents satisfy the predicate on B, we find that the plan chosen by MongoDB (the index scan) can execute over 400\% slower than the best plan for that query. The average slowdown of MongoDB's choice compared to the best possible is 140\%.

%Figure \ref{fig:casec4}, \ref{fig:casec5} and \ref{fig:casec6}  visualize query plans selected by MongoDB V2, the optimal query plans, and impact factors in this scenario. The experiment results verified our observation in section \ref{sec: eti}. That is, \approachName has the advantage that it is insensitive to the distribution of data.

% \begin{figure}[h]
%     \centering
%     \includegraphics[width=0.4\linewidth]{images/body/uniform_dist_practical_5171_10-23-2019_11_41_32.png}
%     \caption{Visualization of query plans selected by MongoDB V2: case 5}
%     \label{fig:vs1}
% \end{figure}

% \begin{figure}[h]
%     \centering
%     \includegraphics[width=0.4\linewidth]{images/body/uniform_dist_actual_5171_10-23-2019_11_41_32.png}
%     \caption{Visualization of the optimal query plans: case 5}
%     \label{fig:vs2}
% \end{figure}

%\begin{figure}[h]
%    \centering
%    \includegraphics[width=0.7\linewidth]{images/body/uniform_dist_practical_5171_10-23-2019_11_41_32.png}
%    \caption{Visualization of query plans selected by MongoDB V2: case 5}
%    \label{fig:vs1}
%
%    \vspace{10pt}
%    \centering
%    \includegraphics[width=0.7\linewidth]{images/body/uniform_dist_actual_5171_10-23-2019_11_41_32.png}
%    \caption{Visualization of the optimal query plans: case 5}
%    \label{fig:vs2}
%
%    \vspace{10pt}
%    \centering
%    \includegraphics[width=0.65\linewidth]{images/body/uniform_dist_error_5171_11-09-2019_15_07_38.png}
%    \caption{Visualization of the impact factor: case 5}
%    \label{fig:vs3}
%\end{figure}

\vspace*{-0.5\baselineskip}
\subsection{Covering Index Scenario}
In our third experiment, we run our usual conjunctive queries over two attributes against a physical database design where, in addition to an index on each attribute, there is also a covering index on the combination (A,B) in that order. The power of the covering index is that query execution often does not need to access the document at all, as the information in the index is enough to determine whether or not the document matches both ranges in the predicate. 

Figure~\ref{fig:coveringindex-evaluation} shows the results of this third experiment. We again first visualize the query plans chosen by MongoDB's \approachName optimizer (Figure~\ref{fig:mongo-coveringindex-choices}), the middle plot shows the optimal plan (Figure~\ref{fig:mongo-coveringindex-optimal}), and the third plot is a visualization of the performance impact of MongoDB's suboptimal plan choices 
(Figure~\ref{fig:mongo-coveringindex-perfimpact}).

% Michael: exp1, comprehensive_summary_accuracy\=54.04_impact_factor\=1.19566.png
From the color distribution in Figure~\ref{fig:mongo-coveringindex-optimal}, we can see that an execution plan using the covering index would indeed be fastest in many cases and that the plan space where a full collection scan is best is now notably smaller than in the previous two scenarios. Furthermore, Figure~\ref{fig:mongo-coveringindex-choices} shows that MongoDB's \approachName optimizer correctly always favors the covering index on (A,B) over one using an index just on A. But again we see that \approachName does not choose the collection scan even in cases where it is best. We see substantial inaccuracy and many cases where the performance is markedly worse than what the true optimal plan would deliver. Although the overall accuracy of \approachName is the best of the three scenarios with about 54\%, the average query speed is still slower than it could be, by 20\%.

\begin{comment}
\subsection{Performance with Varying Data Distributions}
Our experiments have so far always queried two numerical attributes with a uniform data distribution. We also did some extensive experiments with varying the data distributions of each attribute; for example, each field A and B could have uniform, normal, or zipfian distributions. However, we did not find any different behavior of MongoDB's query optimizer in these experiments from that shown so far for a uniform data distribution. The reason is probably that MongoDB chooses its best plan as soon as the first candidate plan produces the 100th result row, which is too early to show any significant difference between the varying data distributions. For space reasons, we omit the details of these runs from this paper and refer the interested reader to our forthcoming technical report. 
\end{comment}

\subsection{Evaluation Summary}
Overall, the experiments in this section demonstrate that while \approachName is a viable approach to query optimization with many good plan choices, it also suffers from what we call ``preference bias'' choosing index scans rather than full collections scans, even when the collection scan would perform much faster.  We explore reasons for this, through a closer look into MongoDB's query optimizer in the next section.
\section{Why does MongoDB's \approachName Optimizer Avoid Collection Scans?}
\label{sec:rootcauseanalysis}
% Nevertheless, we observed that the change we made still not sufficient to always generate optimal query plans. The reasoning behind the preference bias is unknown. 
As we just saw in Section \ref{sec:evaluation}, MongoDB's \approachName optimizer doesn't choose collection scan plans, even for queries when it would run substantially faster than using an index. In this section, we take a closer look at the optimizer code to identify the reasons for this preference bias issue in MongoDB.

Inspection of the query optimizer code revealed a surprising design choice: unless forced to do so, \relname does not include a collection scan among the list of candidate plans to be run in \approachName, if an index is available to satisfy a query.  In more detail,  \texttt{src/\-mongo/\-db/\-query/\-query\_planner.cpp} \footnote{\href{https://github.com/mongodb/mongo/blob/r7.0.1/src/mongo/db/query/query_planner.cpp\#L1607}{https://github.com/mongodb/mongo/blob/r7.0.1/src/mongo/db/query/ query\_planner.cpp\#L1607}} contains this check before a collection scan is considered:

\begin{verbatim}
   if (possibleToCollscan &&
      (collscanRequested || 
       collScanRequired ...))
\end{verbatim}

The variable \texttt{possible\-To\-Collscan} indicates whether a collection scan is possible (database administrators can disable collection scans or the query can include a hint that requires the use of an index); \texttt{collscanRequested} indicates that an explicit query hint has specified that a collection scan should be used; and \texttt{coll\-Scan\-Required} is true only if there is no matching index.  In other words, if a matching index exists, a collection scan must be explicitly requested for MongoDB to consider it.  If the collection scan is not participating, it cannot win the race.

\begin{figure}[htb]
    \centering
    \includegraphics[width=0.9\columnwidth]{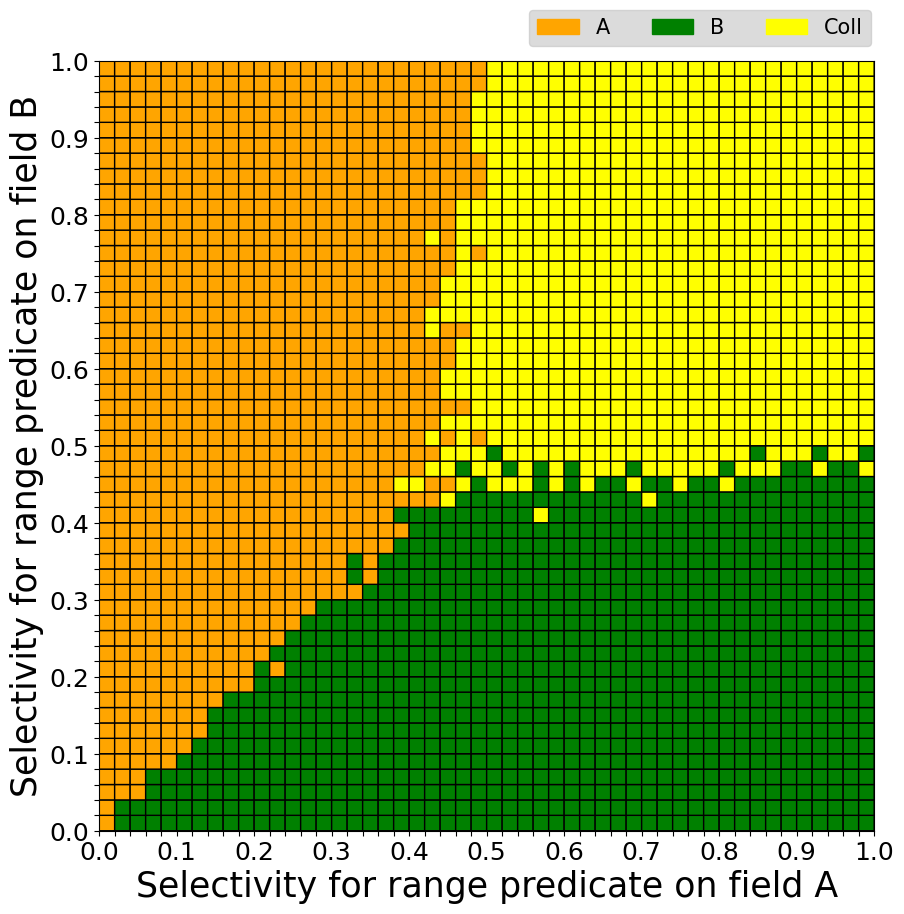}
    \caption{Chosen plans by MongoDB+COLLSCAN.}
    \label{fig:mongo-v1-bothindexed-choices}
\end{figure}

\subsection{Forcing consideration of the collection scan}
However, is this all there is to the issue? We modified the source code to produce a variant we call MongoDB+COLLSCAN that simply always adds a COLLSCAN plan to the set of candidate plans which MongoDB's \approachName optimizer tries out. We repeated our experiments with this variant DBMS and found that it sometimes chooses collection scans as seen in Figure~\ref{fig:mongo-v1-bothindexed-choices}. This is in contrast to  unmodified \relname, which never chooses a collection scan, as shown in Figure~\ref{fig:bothindexed-evaluation}~(a). When we forced COLLSCAN to be considered in the \approachName race, it often does not win, even for queries where the collection scan plan truly runs substantially faster than an alternative plan with index scan.

\begin{figure*}[tb]
%\newlength\plotheight%
\setlength\plotheight{\heightof{\includegraphics[width=0.3\textwidth]{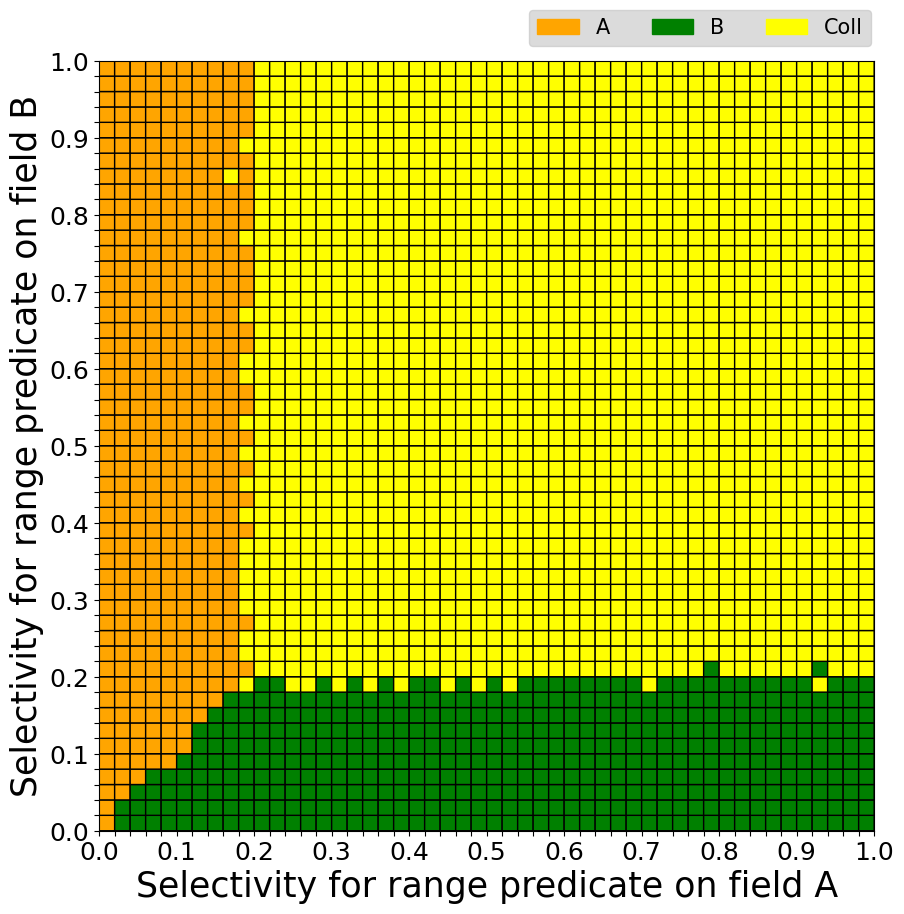}}}
    \begin{subfigure}[t]{0.3\linewidth}
    \includegraphics[height=\plotheight]{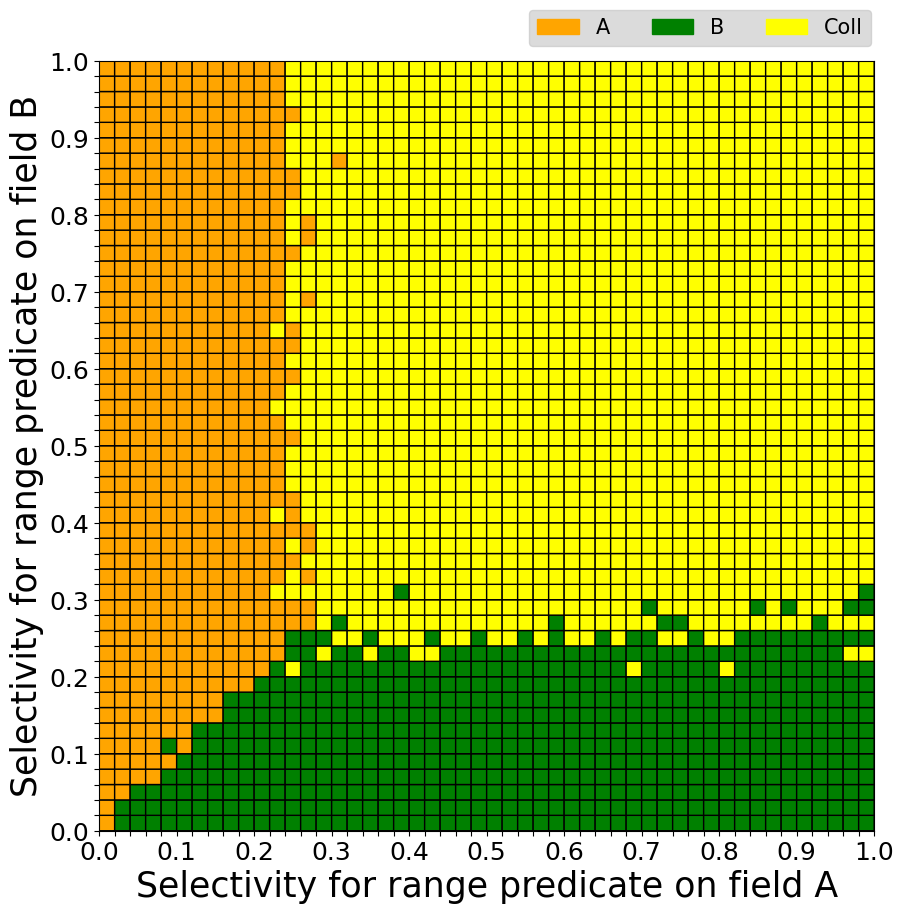}
    \caption{Plans chosen by MongoDB\_MOD.}
    \label{fig:mongo-v2-choices}
    \end{subfigure}
    \hfill
    \begin{subfigure}[t]{0.3\linewidth}
    \includegraphics[height=\plotheight]{images/results-without-covering-index/v\mdbver-with-coll-with-fix/comprehensive_practical_winner.png}
    \caption{Optimal Plan Choices.}
    \label{fig:mongo-v2-optimal}
    \end{subfigure}
    \hfill
    \begin{subfigure}[t]{0.3\linewidth}
    \includegraphics[height=\plotheight-1em]{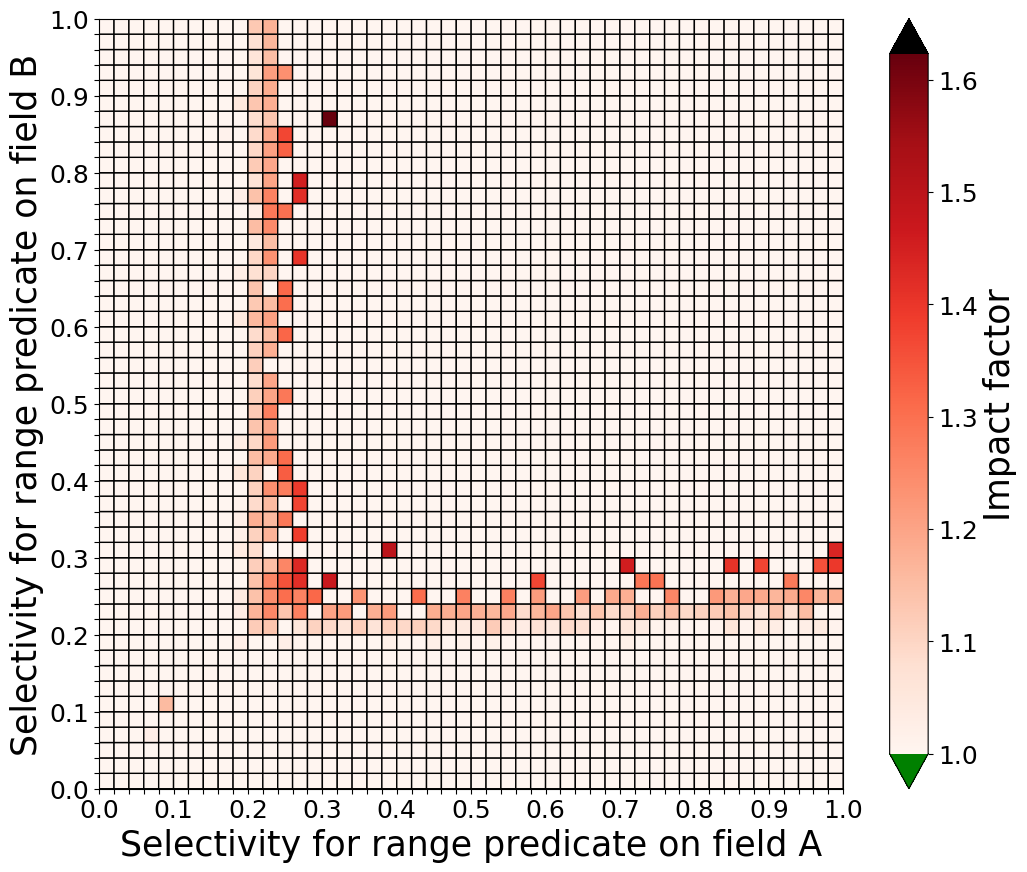}
    \caption{Performance Impact of MongoDB\_MOD's choices.}
    \label{fig:mongo-v2-perfimpact}
    \end{subfigure}
    \caption{Effectiveness of modified \approachName query optimizer of MongoDB\_MOD (dual index scenario).}
    \label{fig:mongo-v2-evaluation}
\end{figure*}

\subsection{Overrated Index Scan}

To further explore the cause of the preference bias that overrates index scans compared to collection scans, we look in detail at the query execution log showing the activity of MongoDB+COLLSCAN during the optimization race for a query with selectivity 0.3 in each attribute. Comparing Figures~\ref{fig:mongo-v1-bothindexed-choices} and \ref{fig:mongo-bothindexed-optimal}, we see that for this query, the optimizer chooses to use an index even when it considers the truly superior collection scan. As mentioned in Section~\ref{sec:background}, the \approachName approach assigns a score to summarize the performance of each query plan at the end of the race (and then the query plan with the highest score will be chosen). The formula considers the $productivity$ of each query plan, where productivity is based on the ratio of the result documents produced to the units of work performed during the race. Note that these work units represent logical steps, not actual measured runtimes.

We discovered that when determining the work units of an index scan, the MongoDB implementation of the \approachName approach ignores the cost of fetching index documents; the optimizer in \relname treats the index retrieval work and the document retrieval work together as a single unit of work (i.e., the same amount of work required by a collection scan looking at the its next document). Therefore, the work unit would likely take more real time for the index scan, and so the productivity of an index scan is overrated. This implementation detail in MongoDB's query optimizer code hides the true advantage of a collection scan in many cases.
\begin{comment}
We suggest that it would be appropriate for MongoDB to adjust the way work unit is measured in the race, to correct this.
\end{comment}

%%%%%%%%%%%%%%%%%%%%%%%%%%%%%%%%%%%%%%%%%%%%
%     impact on the covering index case    % 
%%%%%%%%%%%%%%%%%%%%%%%%%%%%%%%%%%%%%%%%%%%%
\begin{figure*}[tb]
%\newlength\plotheight%
\setlength\plotheight{\heightof{\includegraphics[width=0.3\textwidth]{images/results-without-covering-index/v\mdbver-with-coll-with-fix/comprehensive_practical_winner.png}}}
    \begin{subfigure}[t]{0.3\linewidth}
    \includegraphics[height=\plotheight]{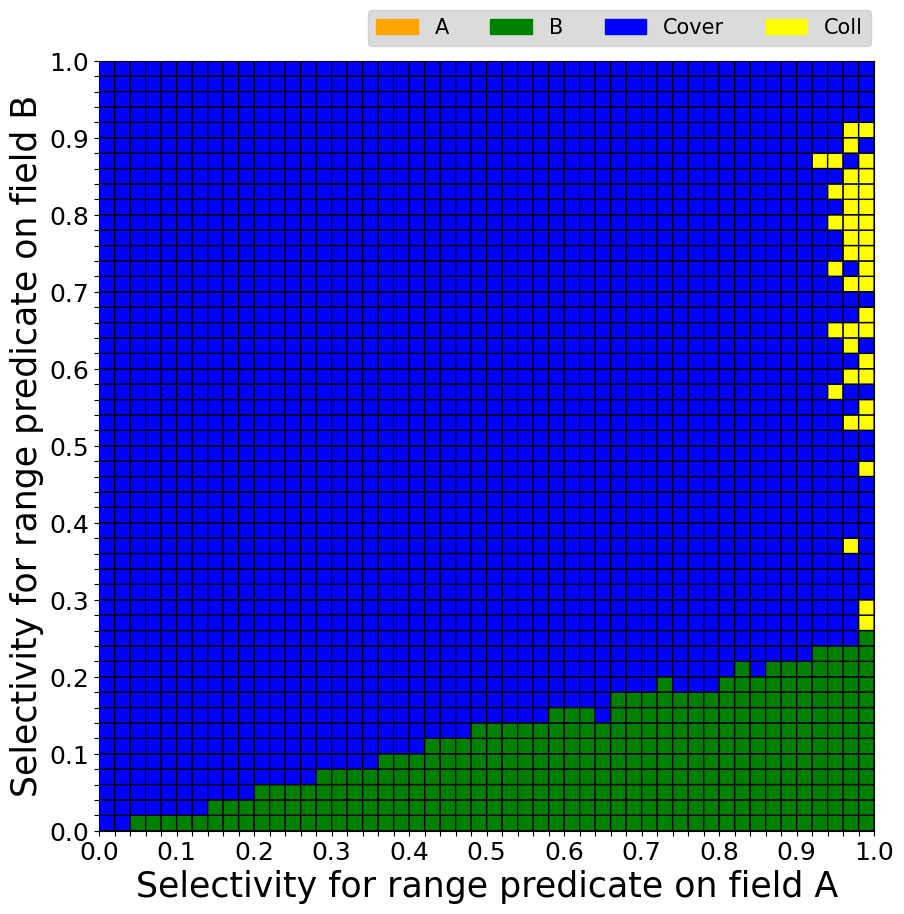}
    \caption{Plans chosen by MongoDB\_MOD.}
    \label{fig:mongo-v2-coveringindex-choices}
    \end{subfigure}
    \hfill
    \begin{subfigure}[t]{0.3\linewidth}
    \includegraphics[height=\plotheight]{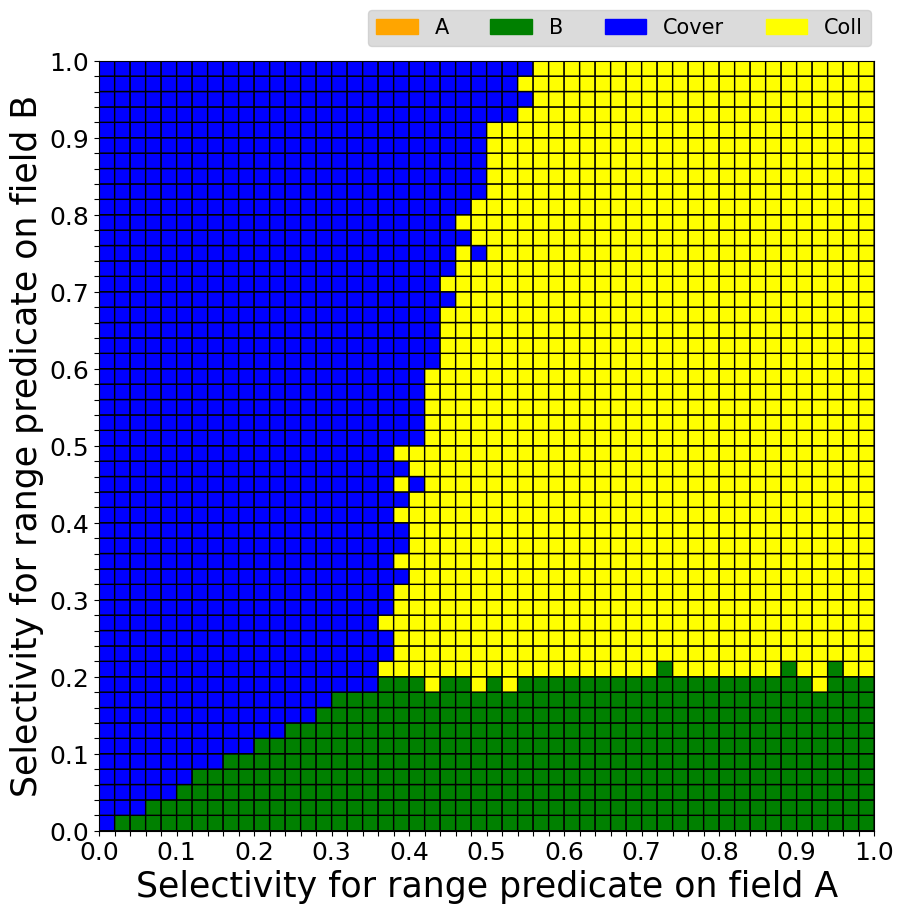}
    \caption{Optimal Plan Choices.}
    \label{fig:mongo-v2-coveringindex-optimal}
    \end{subfigure}
    \hfill
    \begin{subfigure}[t]{0.3\linewidth}
    \includegraphics[height=\plotheight-1em]{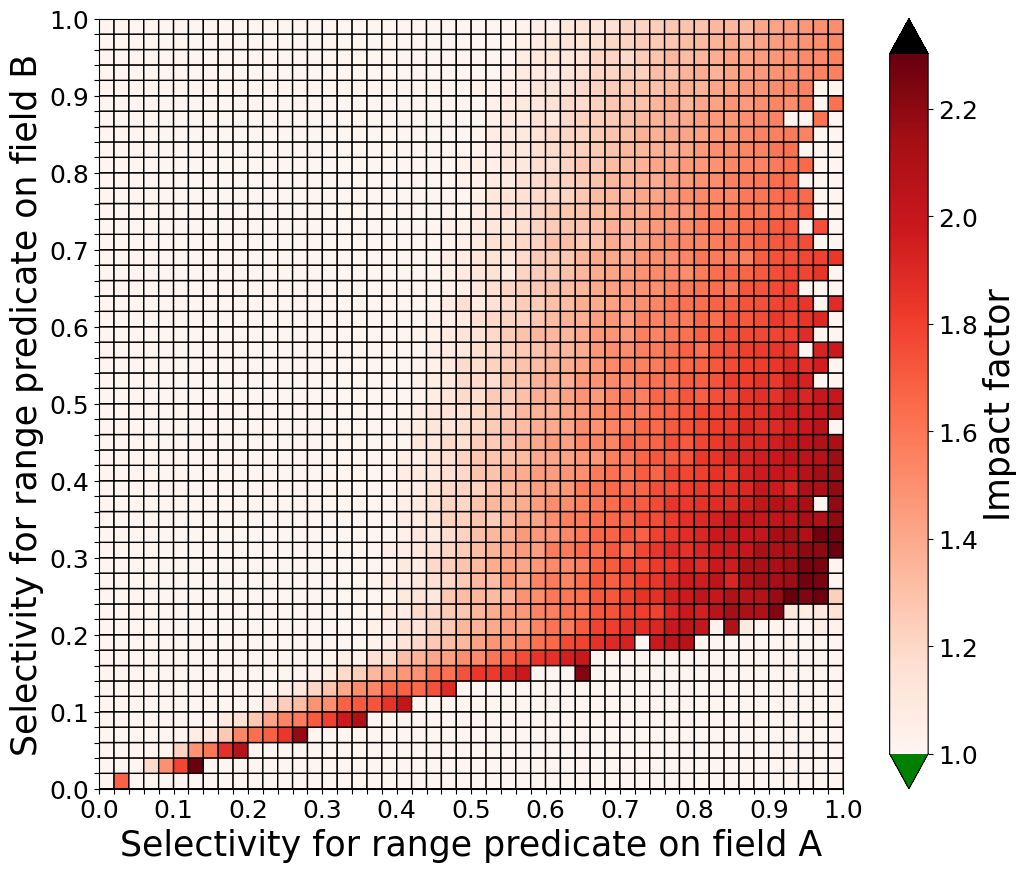}
    \caption{Performance Impact of MongoDB\_MOD's choices.}
    \label{fig:mongo-v2-coveringindex-perfimpact}
    \end{subfigure}
    \caption{Effectiveness of modified \approachName query optimizer of MongoDB\_MOD (covering index scenario).}
    \label{fig:mongo-v2-evaluation-coveringindex}
\end{figure*}

\subsection{Adjusting productivity score of index scan}
We showed above that the optimizer ascribes too much productivity to the index scan, because it treats as a work unit the combination of retrieving the index entry and retrieving the document the index points to. To understand better the impact of this, we made another small modification to the optimizer code. When the plan contains a FETCH (that is, when there is a lookup through the index) we halved the productivity score calculated. This is overly simplistic for complex query plans, but its a reasonable shortcut for the simple queries in these experiments. We call the variant system with forced consideration of COLLSCAN, and also with the adjusted productivity score as described, MongoDB\_MOD.  This system is evaluated for the physical design with uniform data distribution and two indexes, in Figures~\ref{fig:mongo-v2-choices}--\ref{fig:mongo-v2-perfimpact}. We see that while the modified score is not a perfect adjustment, it makes the right decision in many of the queries, and the chosen plan is never much worse than the best possible. We remark that, in theory, the optimal plan choice should be exactly the same for this system as for unmodified \relname since the only change is in the optimizer rather than in query plan execution (that is, Fig~\ref{fig:mongo-v2-optimal} should be the same as Fig~\ref{fig:mongo-bothindexed-optimal}. This is not exactly the case because each figure is produced from measurements of the running time of the plans, and there is some experimental variation from run to run; however, the differences are only occasional cells near the region boundary, where two plans have almost the same cost, so which is fastest (by a tiny amount) can vary between runs.

% Michael exp3-2, Accuracy: 90.88% Impact factor: 1.0145401778590586

The overall accuracy of the query optimizer of this modified version of MongoDB is now 91\% (up from just 34\% measured in Section~\ref{sec:evaluation_bothindexed}), and the average performance impact of the remaining suboptimal plan choices is 1.5\%. 

% Michael exp3, comprehensive_summary_accuracy\=52.48_impact_factor\=1.21331.png
The suggested adjustment of the index scan score also helps somewhat in the scenario with a physical design that includes covering indexes, as shown in Figure~\ref{fig:mongo-v2-evaluation-coveringindex}. If we compare the plan choices of the modified \approachName optimizer in Figure~\ref{fig:mongo-v2-coveringindex-choices} with the ones done by the original \relname optimizer in Figure~\ref{fig:mongo-coveringindex-choices}, we not only see collection scans sometimes being chosen, but also note the much broader use of the covering index, which resembles much closer the optimal case (the few variations near the region boundaries between the optimal cases between Figures~\ref{fig:mongo-v2-coveringindex-optimal} and \ref{fig:mongo-coveringindex-optimal} are due to slight variations in the runtimes between experiments). % SURE? [ur]
Despite these qualitative improvements, the overall accuracy of the modified query optimizer has not changed significantly, at 52.5\%, with an average performance impact of 21\%.

\subsection{Discussion}
We have shown that MongoDB's implementation of \approachName query optimization has a systematic preference bias, avoiding collection scans, and that this can lead to poor choice of execution plan. The outcomes can be improved by forcing consideration of COLLSCAN in the race, and adjusting the productivity score to recognize the extra work done 
%in an index scan 
when both index lookup and then document fetch happen. More work is needed to find more sophisticated ways to score productivity that will deal with complex plans with index lookup on some but not all steps. 
%(and so we need an adjustment that is more nuanced than simply halving the score).

The scale of the workload evaluated here is considerably smaller than that common in real-world settings: we only measured cases where all indexes fit in memory. In reality, companies often have volumes of data stored in MongoDB that are much larger than RAM. This would significantly magnify the true costs of the index scan and thus dramatically increase the negative impact of the preference bias from neglecting the cost of fetching index entries. Therefore, in such cases, we expect that preference bias will become a much more serious issue.

\newpage
\section{Conclusions}
\label{sec:conclusions}

MongoDB uses a unique approach to query optimization, which we call \approachName query optimization. This differs significantly from the traditional cost-based query optimization in that it chooses its execution plans based on an "execution race" with multiple candidate plans. To the best of our knowledge, this is the first paper explaining and evaluating this approach.

We analyzed the effectiveness of MongoDB's \approachName optimizer using experiments that consider a set of queries, which adjust parameter values to vary the selectivity. For each query, we find which plan the MongoDB optimizer chooses, which plan is actually the fastest in execution, and we see how much worse the chosen plan is compared to the optimal one. Each of these aspects is displayed on a grid of cells. These displays provide two plan diagrams~\cite{reddy2005analyzing} and an innovative heatmap visualization of the impact on performance of optimizer choices.  This visualization makes it easy to visually identify areas where query planning can be improved. We also provide two summary numbers for accuracy and average performance impact, of the optimizer's choice of plan compared to that which is truly the best.

We showed that \relname has a query plan preference bias: the current implementation of \approachName does not consider a collection scan among the candidate plans unless the client requests it or there is no alternative. Furthermore, we showed that MongoDB calculation of plan productivity (during the race) underestimates the costs of an index scan where both index entry and document must be accessed. This leads MongoDB to often choose an index-based plan, even when it is forced to compare this to a collection scan. These issues mean that \approachName often chooses a plan that can be substantially slower than an optimal collection scan for some queries that retrieve many of the documents.

This paper offers the beginning of a detailed study of MongoDB's query optimizer.
%So far we have only considered simple queries that run over a collection of documents with two atomic fields, and use range predicates on each field.
In future work, we will examine how the optimizer works on more complicated document schemas and more complicated queries. We also intend to explore more sophisticated ways to improve the productivity scoring, so it comes closer to the true costs of the work done in a plan during the race.
\normalsize
\bibliography{_main}

%\appendix
%\input{figures-with-explanations.tex}
%\input{effect-of-caching.tex}

%\input{7-biographies}

\end{document}